\documentclass[fleqn,usenatbib]{mnras}

\usepackage{newtxtext,newtxmath}

\usepackage[T1]{fontenc}

\DeclareRobustCommand{\VAN}[3]{#2}
\let\VANthebibliography\thebibliography
\def\thebibliography{\DeclareRobustCommand{\VAN}[3]{##3}\VANthebibliography}

\usepackage{graphicx}	
\usepackage{amsmath}	
\usepackage{threeparttable}

\title[Phase-resolved spectroscopy of GX 301--2]{Evidence for a nearly orthogonal rotator in GX 301--2 with phase-resolved cyclotron resonant scattering features }

\author[X. Chen et al.]{
Xiao Chen,$^{1,2}$
Yuanze Ding,$^{1,2}$
Wei Wang,$^{1,2}$\thanks{E-mail: wangwei2017@whu.edu.cn}
Osamu Nishimura,$^{3}$
Qi Liu,$^{1,2}$
Shuang-Nan Zhang,$^{4}$
\newauthor
Mingyu Ge,$^{4}$
Fangjun Lu,$^{4}$
Jinlu Qu,$^{4}$
Liming Song$^{4}$
and Shu Zhang$^{4}$
\\g
$^{1}$Department of Astronomy, School of Physics and Technology, Wuhan University, Wuhan, 430072, China\\
$^{2}$WHU-NAOC Joint Center for Astronomy, Wuhan University, Wuhan, 430072, China\\
$^{3}$National Institute of Technology, Nagano College, 716 Tokuma, Nagano City, 381-8550, Japan\\
$^{4}$Key Laboratory of Particle Astrophysics, Institute of High Energy Physics, Chinese Academy of Sciences, Beijing, 100049, China
}

\date{Accepted XXX. Received YYY; in original form ZZZ}

\pubyear{2024}

\begin{document}
\label{firstpage}
\pagerange{\pageref{firstpage}--\pageref{lastpage}}
\maketitle

\begin{abstract}
Cyclotron resonant scattering features (CRSFs) are the absorption features in the X-ray spectra of strongly magnetized accretion neutron stars (NSs), which are probably the most reliable probe to the surface magnetic fields of NSs. The high mass X-ray binary GX 301--2 exhibits a very wide, variable and complicated CRSF in the average spectra, which should be two absorption lines based on NuStar and Insight-HXMT observations. With the Insight-HXMT frequent observations, we performed the phase-resolved spectroscopy and confirmed two cyclotron absorption lines in the phase-resolved spectra, with their centroid energy ratio $\sim 1.6-1.7$ in the super-critical luminosity case. A major hindrance in understanding those CRSFs is the very poorly constrained magnetic inclination angle, which is also a fundamental property of a NS and key to understanding the emission characteristics of a pulsar. Comparing the phase-resolved CRSF with simulated X-ray spectra, the magnetic inclination angle is found to be $\gtrsim 70^{\circ}$, i.e., nearly orthogonal between the NS's spin and magnetic axies. The implications of an orthogonal rotator and magnetic structure evolution in the accreting X-ray binary are also discussed.
\end{abstract}




\section{Introduction}

Neutron stars (NS) are ideal astrophysical laboratories for testing theories for dense materials, gravity and strong magnetic field physics \citep{Feryal2016}. Rotating NS, namely pulsar, emits radio, X-ray and multiwavelength pulsed radiation, and provides a tool to probe the NS characteristics \citep{Manchester2004}. The magnetic inclination as an angle between the magnetic and spin axis is a key parameter for understanding the radiation mechanism, but is not well understood yet \citep{Beskin1993ppm,Novoselov2020MNRAS}, especially for the accretion X-ray pulsars, which we still have no clear observational information \citep{Biryukov2021MNRAS}. Cyclotron resonant scattering features are the absorption features in strongly magnetized accretion neutron stars \citep{Truemper1978ApJ,Staubert2019A&A}, which can show the accretion processes near the magnetic poles of neutron stars and the magnetic field geometry \citep{Lamb1973ApJ,Harding1991ApJ,Schonherr2007A&A,Nishimura2011ApJ,Mushtukov2022arXiv}.

GX 301--2 contains a neutron star of $M_{\rm NS}\simeq 1.85\pm 0.60 M_{\odot}$ \citep{Kaper2006A&A} orbiting a B1 hypergiant star Wray 977 with a 41.5\,d eccentric orbit \citep{White1984ApJ} at a distance of $\sim 3.5$ kpc \citep{Bailer-Jones2018AJ}. GX 301--2 has a slowly rotating pulsar ($P_{\rm spin}\simeq 680\,\rm s$) \citep{Ding2021MNRAS,monkkonen2020} compared to the majority of other accreting X-ray pulsars, and shows strong orbital X-ray variability \citep{Sato1986ApJ,Nabizadeh2019A&A,Abarr2020ApJ}. The regular Type-1 outburst occurs shortly before every periastron passage of the NS \citep{Leahy1991MNRAS,Leahy2002A&A,Leahy2008MNRAS}, resulting in the intense X-ray flares with luminosity higher than $10^{37}\rm \,erg~s^{-1}$ \citep{Islam2014MNRAS}. Recently, the Fe~K$\alpha$ time-lag observations in GX 301--2 suggested accretion from a tidal stream rather than quasi-isotropic stellar winds \citep{Zheng2020MNRAS}, which is supported by the formation of transient disk during the spin-up event \citep{Liu2021MNRAS}.

The cyclotron resonant scattering feature of GX 301--2 around 40 keV was first discovered in \textsl{Ginga} observations \citep{Makishima1992}, and then confirmed by other missions \citep{Kreykenbohm2004A&A,LaBarbera2005A&A,Doroshenko2010A&A}. However, the absorption feature is very broad, and its centroid energy varies wildly from $\sim 30 -55$ keV in different observations \citep{Suchy2012ApJ}, which makes the physical origin as a longstanding puzzle \citep{Ding2021MNRAS}. With high resolution spectroscopy by \textsl{NuSTAR}, it has been argued that the reported broad feature is in fact from two separate line spawning regions located at the NS surface and 1.4\,km above, generating two cyclotron absorption lines at $E_1\sim 35-40$ keV and $E_2\sim 50-55$ keV respectively \citep{Furst2018A&A}. The recent \textsl{Insight}-HXMT data confirmed two absorption lines in the average spectra, with a line ratio of $\sim 1.7$ based on the multiple observations \citep{Ding2021MNRAS}.

GX 301--2 has been observed by the \textsl{Insight}-HXMT for multiple times from August 2017 to December of 2020, showing the variable X-ray light curves over time and orbital phases \citep{Ding2021MNRAS}. Phase-average spectroscopy based on \textsl{Insight}-HXMT shows that the centroid energies of both two CRSFs vary with X-ray luminosity, discovering the presence of critical luminosity and accretion state transition in GX 301--2 \citep{Ding2021MNRAS}.

Modeling the structure of accretion column in accreting X-ray NS is of fundamental importance as it would answer how the materials are transferred into radiation around the NS, and could potentially unveil the inner structure and evolution of NS. However, such a task is till now a formidable problem due to the coupling of radiation transfer and magnetic fluid dynamical processes in the extremely strong magnetic field in NS atmosphere. There have been several attempts to model the radiation from NS accretion column \citep{Schonherr2007A&A,Nishimura2011ApJ,nishimura2015,Schwarm2017A&A,Nishimura2008ApJ,Nishimura2014ApJ}. To reduce the intractable complexity in solving the time-dependent radiation transfer and magnetic hydro-dynamical equations, a common strategy in these works is to adopt an analytic super-critical accretion column solution, which assumes a magnetically confined ideal fluid impact on the NS magnetic polar, producing radiation dominated shock \citep{Nishimura2008ApJ,Becker1998ApJ,Arons1992ApJ,Basko1976MNRAS}. The velocity distribution along the vertical axis results in the Doppler boosting on the generated CRSFs, thus producing quite different absorption structure when changing the viewing angle \citep{nishimura2015,nishimura2022}.

In this work, the \textsl{Insight}-HXMT observation and the data reduction of the phase-resolved spectrum are described in Section 2. Then, the simulation process is presented in Section 3, including the spectral modelling and the generation of the simulated phase-resolved spectra. In Section 4, the results of both observational and simulated phase-resolved spectroscopic data are presented. The dependency of CRSF properties on pulse phases and magnetic configuration, which provides a unique method to constrain the magnetic inclination of NS, is discussed in Section 5. Finally, we summarize our results in Section 6.

\section{Observation and data reduction}\label{sec:method}

\subsection{\textsl{Insight}-HXMT}

\textsl{Insight}-HXMT has three main payloads: the Low Energy X-ray Telescope (LE, \citealt{Chen2020SCPMA}), the Medium Energy X-ray Telescope (ME, \citealt{Cao2020SCPMA}), and  the High Energy X-ray Telescope (HE, \citealt{Liu2020SCPMA}). HE contains 18 cylindrical NaI(Tl) / CsI(Na) detectors with a total detection area of 5000 cm$^2$ in the energy range of 20--250 keV. The time resolution is 25 $\mu$s, and its energy resolution at 60 keV is 19\% \citep{Zhang2020SCPMA}. ME is composed of 1728 Si-PIN detectors with a total detection area of 952 cm$^2$ covering the range of 5--30 keV. Its time resolution is 280 $\mu$s, and the energy resolution at 20 keV is 14\%. LE uses Swept Charge Device (SCD) with a total detection area of 384 cm$^2$ covering 1--15 keV. It has a time resolution of 1 ms and the energy resolution at 6 keV is 2.5\%. The typical Field of Views (FoVs) for the three detectors are $1.1^{\circ}\times5.7^{\circ}$ (HE), $1^{\circ}\times4^{\circ}$ (ME) and $1.6^{\circ}\times6^{\circ}$ (LE) \citep{Zhang2020SCPMA}. \textsl{Insight}-HXMT has carried out a series of performance verification tests by observing blank sky, standard sources, showing good calibration state and estimation of the instrumental background \citep{Li2020JHEAp}.

\textsl{Insight}-HXMT satellite conduct intensive recurring pointing observations on GX 301--2 between August~3$^{rd}$,~2017 and June~3$^{rd}$,~2020 (from MJD 57968 to 59003), with typical net exposure time in each pointing $\sim 3\,$ks. These observations cover almost all orbital phases of the binary orbit. The observations (information of OBSIDs) used in this paper are the same as those in \cite{Ding2021MNRAS}, and they are listed in Table~\ref{tab:obsdata}, along with the observation date, exposure, count rate, pulse period, orbital phase, and luminosity. In the following science analysis, we filtered data with the recommended settings in HXMT Data Reduction Guide: (1) pointing offset angle $< 0.1^\circ$; (2) pointing direction above Earth $> 10^\circ$; (3) geomagnetic cut-off rigidity value $> 8\,$GeV; (4) time since SAA passage $> 300\,$s and time to next SAA passage $> 300\,$s; (5) for LE observations, pointing direction above bright Earth $> 30^\circ$. The \textsl{Insight}-HXMT Data Analysis (HXMTDAS) V2.04 was used to produce high level products. We made the barycentric correction of the data using the tool \textsc{hxbary} in HXMTDAS. Well-calibrated energy bands of LE, ME and HE were used in the following spectra analysis ($3-8.5\,$keV, $10-30\,$keV and $30-70\,$keV, respectively, \citealt{Li2020JHEAp}). ME data in $21-24\,$keV was ignored due to the significant increase of calibration uncertainties in this energy range.

\begin{table*}
	\centering
	\caption{List of observations analyzed in this paper. The observation date, start MJD, HE exposure, count rates of three detectors, pulse period, orbital phase, and luminosity are also listed.}
	\label{tab:obsdata}
	\begin{tabular}{cccccccccc}
\hline
\hline
OBSID	&	 Start Date	&	 Start MJD	&	HE Exposure	&	HE rate		&		LE rate		&		ME rate		&		Pulse Period		&		Orbital Phase	&	Luminosity					\\
 & (YYYYMMDD) & & (s) & (cts/s) & (cts/s) & (cts/s) & (s) & & ($\times10^{37}$ erg/s) \\
\hline
P010130900103	&	20170803	&	57968.54 	&	3506	& $	444.4 	\pm	0.4 	$ & $	31.7 	\pm	0.1 	$ & $	213.0 	\pm	0.3 	$ & $	685.3 	\pm	0.1 	$ &	0.959 	& $	4.216 	\pm	0.012 	$		\\
P010130900104	&	20170803	&	57968.68 	&	2537	& $	344.0 	\pm	0.4 	$ & $	18.7 	\pm	0.1 	$ & $	99.1 	\pm	0.2 	$ & $	685.3 	\pm	0.1 	$ &	0.963 	& $	1.711 	\pm	0.010 	$		\\
P010130900107	&	20170804	&	57969.07 	&	5309	& $	379.6 	\pm	0.3 	$ & $	23.3 	\pm	0.1 	$ & $	108.6 	\pm	0.2 	$ & $	685.3 	\pm	0.1 	$ &	0.972 	& $	1.848 	\pm	0.007 	$		\\
P010130900401	&	20170804	&	57969.51 	&	3849	& $	303.3 	\pm	0.3 	$ & $	18.5 	\pm	0.1 	$ & $	48.8 	\pm	0.1 	$ & $	685.3 	\pm	0.1 	$ &	0.983 	& $	0.660 	\pm	0.006 	$		\\
P010130900402	&	20170804	&	57969.67 	&	2422	& $	401.6 	\pm	0.4 	$ & $	40.8 	\pm	0.2 	$ & $	157.3 	\pm	0.3 	$ & $	685.3 	\pm	0.1 	$ &	0.987 	& $	2.745 	\pm	0.010 	$		\\
P010130900503	&	20180103	&	58121.38 	&	2686	& $	316.7 	\pm	0.4 	$ & $	16.7 	\pm	0.1 	$ & $	23.3 	\pm	0.1 	$ & $	682.0 	\pm	2.0 	$ &	0.644 	& $	0.253 	\pm	0.009 	$		\\
P010130900702	&	20180119	&	58137.65 	&	2883	& $	322.4 	\pm	0.4 	$ & $	16.5 	\pm	0.1 	$ & $	43.3 	\pm	0.2 	$ & $	683.7 	\pm	0.1 	$ &	0.036 	& $	0.625 	\pm	0.008 	$		\\
P010130900703	&	20180119	&	58137.80 	&	2096	& $	340.4 	\pm	0.4 	$ & $	15.0 	\pm	0.1 	$ & $	34.5 	\pm	0.1 	$ & $	683.7 	\pm	0.1 	$ &	0.040 	& $	0.741 	\pm	0.011 	$		\\
P010130900704	&	20180119	&	58138.01 	&	738	& $	385.9 	\pm	0.8 	$ & $	14.7 	\pm	0.1 	$ & $	34.9 	\pm	0.1 	$ & $	683.7 	\pm	0.1 	$ &	0.043 	& $	0.564 	\pm	0.013 	$		\\
P010130900705	&	20180120	&	58138.04 	&	3206	& $	353.4 	\pm	0.3 	$ & $	15.2 	\pm	0.1 	$ & $	42.3 	\pm	0.1 	$ & $	683.7 	\pm	0.1 	$ &	0.046 	& $	0.579 	\pm	0.007 	$		\\
P010130900707	&	20180120	&	58138.33 	&	2062	& $	389.2 	\pm	0.5 	$ & $	18.8 	\pm	0.1 	$ & $	75.2 	\pm	0.2 	$ & $	683.7 	\pm	0.1 	$ &	0.052 	& $	1.033 	\pm	0.009 	$		\\
P010130900708	&	20180120	&	58138.44 	&	2584	& $	321.9 	\pm	0.4 	$ & $	14.7 	\pm	0.1 	$ & $	27.0 	\pm	0.1 	$ & $	683.7 	\pm	0.1 	$ &	0.055 	& $	0.335 	\pm	0.007 	$		\\
P010130900710	&	20180120	&	58138.73 	&	2864	& $	319.3 	\pm	0.3 	$ & $	15.9 	\pm	0.1 	$ & $	40.3 	\pm	0.1 	$ & $	683.7 	\pm	0.1 	$ &	0.062 	& $	0.484 	\pm	0.007 	$		\\
P010130900809	&	20180130	&	58148.80 	&	5835	& $	331.7 	\pm	0.3 	$ & $	16.5 	\pm	0.1 	$ & $	24.0 	\pm	0.1 	$ & $	684.0 	\pm	1.7 	$ &	0.305 	& $	0.331 	\pm	0.006 	$		\\
P010130901601	&	20180520	&	58258.27 	&	4102	& $	472.6 	\pm	0.4 	$ & $	30.0 	\pm	0.1 	$ & $	169.3 	\pm	0.3 	$ & $	683.8 	\pm	0.5 	$ &	0.944 	& $	3.549 	\pm	0.011 	$		\\
P010130901602	&	20180520	&	58258.44 	&	3060	& $	413.0 	\pm	0.4 	$ & $	19.7 	\pm	0.1 	$ & $	117.0 	\pm	0.2 	$ & $	683.8 	\pm	0.5 	$ &	0.948 	& $	2.586 	\pm	0.016 	$		\\
P010130902103	&	20190303	&	58545.69 	&	656	& $	455.8 	\pm	0.8 	$ & $	35.4 	\pm	0.1 	$ & $	83.4 	\pm	0.2 	$ & $	672.3 	\pm	0.4 	$ &	0.870 	& $	1.613 	\pm	0.014 	$		\\
P010130902104	&	20190303	&	58545.72 	&	6536	& $	453.1 	\pm	0.3 	$ & $	38.7 	\pm	0.1 	$ & $	113.6 	\pm	0.1 	$ & $	672.3 	\pm	0.4 	$ &	0.873 	& $	1.977 	\pm	0.006 	$		\\
P020101228001	&	20190706	&	58670.44 	&	1350	& $	443.1 	\pm	0.6 	$ & $	22.9 	\pm	0.2 	$ & $	62.4 	\pm	0.2 	$ & $	686.1 	\pm	0.4 	$ &	0.899 	& $	1.154 	\pm	0.012 	$		\\
P020101228004	&	20190706	&	58670.77 	&	2303	& $	437.9 	\pm	0.5 	$ & $	24.3 	\pm	0.1 	$ & $	84.2 	\pm	0.2 	$ & $	686.1 	\pm	0.4 	$ &	0.923 	& $	1.523 	\pm	0.013 	$		\\
P020101228005	&	20190706	&	58670.90 	&	3144	& $	412.1 	\pm	0.4 	$ & $	23.8 	\pm	0.1 	$ & $	73.7 	\pm	0.2 	$ & $	686.1 	\pm	0.4 	$ &	0.923 	& $	1.425 	\pm	0.011 	$		\\
P020101228007	&	20190707	&	58671.19 	&	2961	& $	409.2 	\pm	0.4 	$ & $	31.9 	\pm	0.1 	$ & $	87.3 	\pm	0.2 	$ & $	686.1 	\pm	0.4 	$ &	0.923 	& $	1.507 	\pm	0.009 	$		\\
P020101228401	&	20200115	&	58863.59 	&	3786	& $	344.3 	\pm	0.3 	$ & $	15.7 	\pm	0.1 	$ & $	24.5 	\pm	0.1 	$ & $	670.0 	\pm	2.1 	$ &	0.577 	& $	0.377 	\pm	0.007 	$		\\
P020101228606	&	20200130	&	58878.05 	&	11800	& $	391.0 	\pm	0.2 	$ & $	26.1 	\pm	0.1 	$ & $	51.7 	\pm	0.1 	$ & $	683.4 	\pm	0.4 	$ &	0.915 	& $	0.914 	\pm	0.004 	$		\\
P020101228610	&	20200130	&	58878.82 	&	2390	& $	400.4 	\pm	0.4 	$ & $	20.7 	\pm	0.1 	$ & $	45.2 	\pm	0.1 	$ & $	683.4 	\pm	0.4 	$ &	0.939 	& $	0.994 	\pm	0.012 	$		\\
P020101228701	&	20200214	&	58893.54 	&	3047	& $	366.9 	\pm	0.4 	$ & $	14.3 	\pm	0.1 	$ & $	21.9 	\pm	0.1 	$ & $	669.0 	\pm	5.7 	$ &	0.300 	& $	0.249 	\pm	0.007 	$		\\
P020101228804	&	20200318	&	58926.85 	&	3237	& $	388.3 	\pm	0.4 	$ & $	19.4 	\pm	0.1 	$ & $	34.8 	\pm	0.1 	$ & $	671.0 	\pm	2.3 	$ &	0.096 	& $	0.455 	\pm	0.007 	$		\\
P020101229002	&	20200603	&	59003.47 	&	4891	& $	452.6 	\pm	0.3 	$ & $	33.0 	\pm	0.1 	$ & $	94.9 	\pm	0.2 	$ & $	669.9 	\pm	0.6 	$ &	0.929 	& $	1.559 	\pm	0.007 	$		\\
P020101229003	&	20200603	&	59003.61 	&	1504	& $	445.0 	\pm	0.6 	$ & $	30.3 	\pm	0.1 	$ & $	78.2 	\pm	0.2 	$ & $	669.9 	\pm	0.6 	$ &	0.953 	& $	1.358 	\pm	0.012 	$		\\
\hline
	\end{tabular}
\end{table*}

\subsection{Phase-resolved spectral analysis}

\begin{figure}
    \centering
    \includegraphics[width=0.48\textwidth]{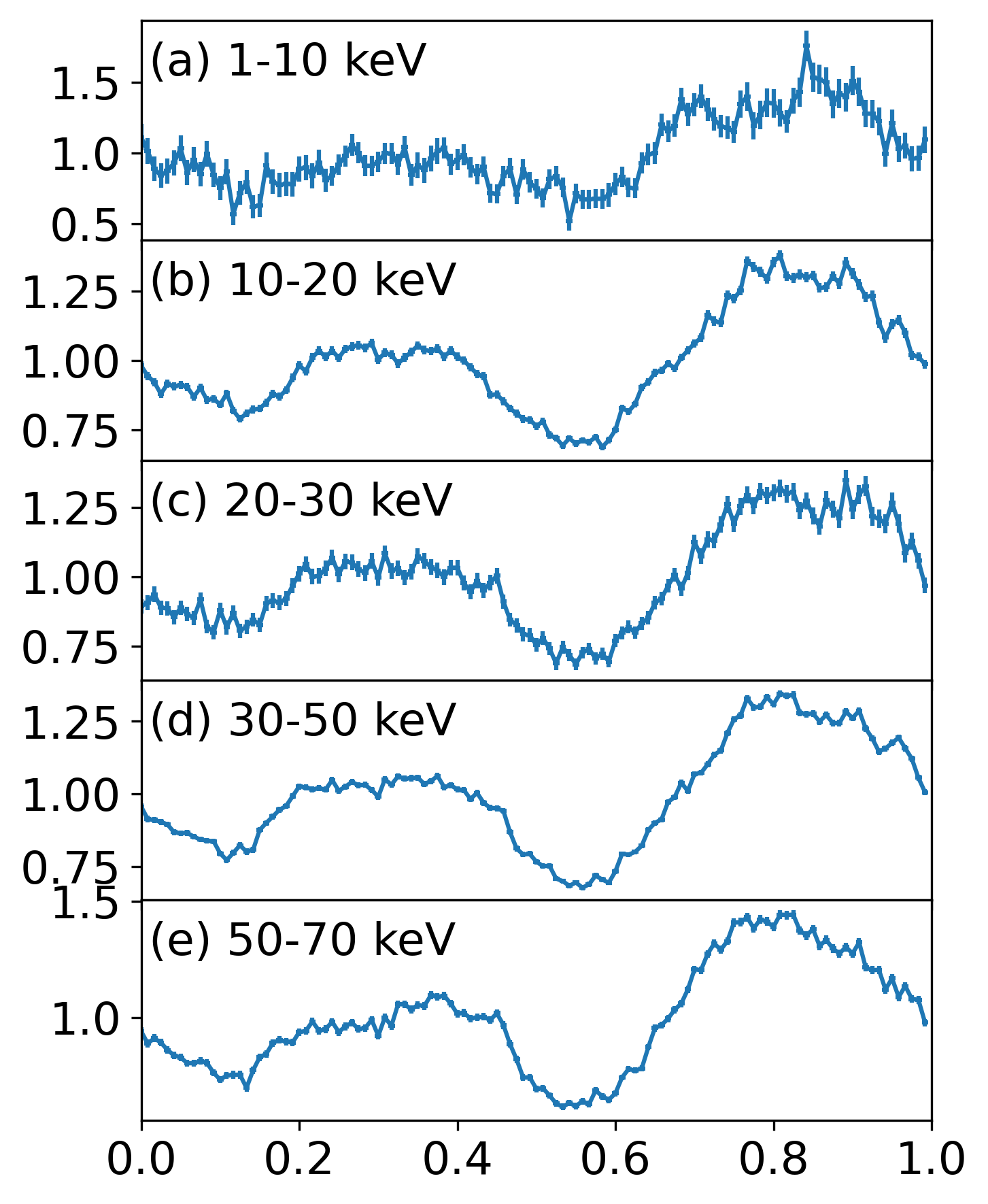}
    \caption{Pulse profiles of P010130900103 in different energy bands.}
    \label{fig:pulseProfile}
\end{figure}

The photon arrival time for binary motion is first corrected using the latest ephemeris provided by \cite{Doroshenko2010A&A}. For each \textsl{Insight}-HXMT observation, we derive the spin period of the neutron star of GX 301--2 using \textit{efsearch} which folds the light curve and finds the best period by searching for the maximum chi-square. The error is then calculated with $\sigma_{f} = \sqrt{2} a \sigma_{tot} / (\sqrt{N} A T)$ provided by \cite{Larsson1996A&AS..117..197L} because of the uneven sampling rate, where $a$ is a simulation constant of 0.469, $N$ is the total data points, $A$ is the sinusoidal amplitude, and $T$ is the time duration. The pulse period and error are shown in Table~\ref{tab:obsdata}. To improve accuracy, pulse periods shown in Table~\ref{tab:obsdata} are determined using combined data where observations taken within two days are merged together. In general, a period of approximately 685 seconds would result in only four pulses being observed during a continuous exposure time of around 3 ks. However, the typical total time span of each \textsl{Insight}-HXMT observation is approximately 9 ks, with the 3 ks exposure time resulting from the selection of Good Time Intervals (GTI). This means that each of our observation covers more than 10 pulses, which is sufficient to obtain a reasonably accurate pulse period. Pulse profiles can then be obtained. Here we show an example of pulse profiles of P010130900103 in different energy bands covering the energy range that used in this paper, see Figure~\ref{fig:pulseProfile}. More detailed pulse information can be found in  \cite{Ding2021MNRAS}. The observed spectra of the accretion X-ray pulsar would vary in different pulse phases, which can be used to infer the accretion process of the emission region and geometric characteristics. In each pointing observation, we calculate the corresponding pulse phase of each corrected time based on the derived pulse period, setting initial phase to be the periastron passage \citep{Doroshenko2010A&A}. By dividing the phase into eight equal width parts (see Section~\ref{sec: obs results}), the observation time can also be divided into the corresponding eight parts, thereby the corresponding phase-resolved spectra can be extracted, with net exposure time of $\sim 150-300\,$s. For each phase-resolved spectrum, we use the HXMT software to regenerate the new background spectrum and response matrix.

Different phenomenological models with a power law modified by cut-off at higher energy can describe the broad band spectrum of GX 301--2 very well \citep{Endo2002ApJ,Furst2018A&A,Ding2021MNRAS}. At lower energy the spectrum gets heavily absorbed due to the presence of a variable neutral absorber \citep{Furst2011A&A}, with the Hydrogen column density in the range of $10^{23}-2\times10^{24}$\,\,$\mathrm{cm^{-2}}$ along the observer's line of sight throughout the orbit, and thus indicating accretion mode from clumpy stellar wind. To explain the low-energy part of the spectrum, previous studies applied a partial covering component in addition to the usual galactic absorption \citep{LaBarbera2005A&A,Mukherjee2004A&A}. In $6-7$\,keV, GX 301--2 shows many fluorescent lines \citep{Furst2011A&A,Ji2021MNRAS} including the bright and prominent Fe~K shell emission lines \citep{Endo2002ApJ}. In this work, the K$\alpha$ and K$\beta$ lines are fitted with Gaussian and are frozen at 6.4 and 7.0\,keV respectively. Their widths are frozen at 10 eV.

The spectral fitting in this paper is conducted with \textsc{Xspec} v12.12.0g \citep{Arnaud1996ASPC}. We conducted a phenomenological fitting similar to \cite{Ding2021MNRAS} with the spectral model in \textsc{Xspec}'s terminology as {\em Constant*TBabs*(TBpcf*CRSFs*Continuum+Gaussian+Gaussian)}, where CRSF is represented using the {\em gabs} model.

\section{Simulation}
\subsection{Spectral Modelling}
\label{sec:simu-specModel}
A prominent accretion mound, inside which is a sinking region, is expected to be formed above $L\sim 1\times 10^{37}\,\mathrm{erg~s^{-1}}$ \citep{Arons1992ApJ} in which the line-forming region is expected to be located around the accretion mound.  In this work, we consider a line-forming region
around an accretion mound with a two-dimensional structure
taking into account gravitational light bending under the assumption of two symmetrically located polar caps like \cite{nishimura2022}. 

\begin{figure}
      \centering
    \resizebox{60mm}{!}{\includegraphics{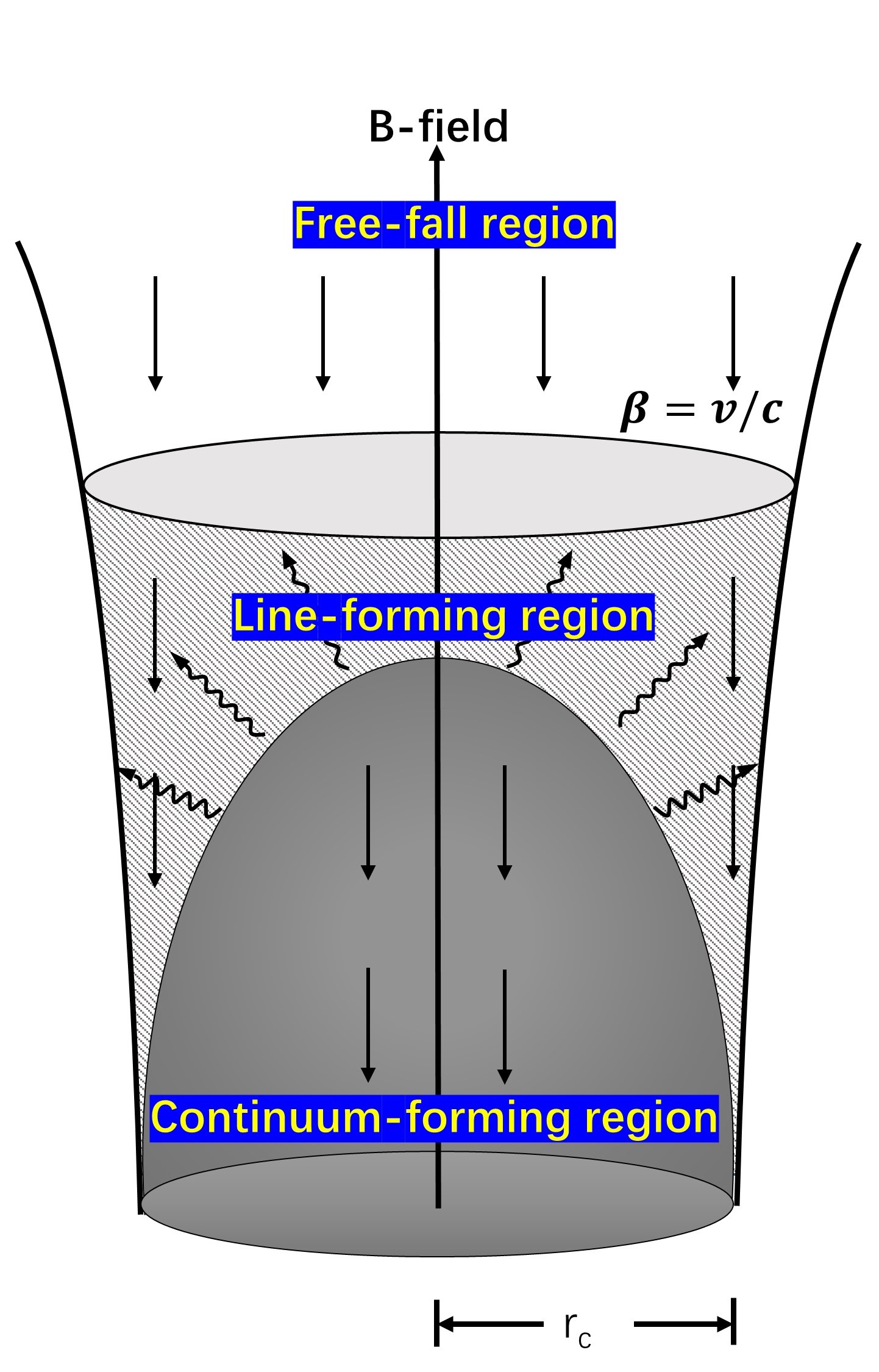}}
  \caption{Schematic of the line-forming region around an accretion mound. The region represented by oblique dot line denotes the line-forming region while the darker shadow region denotes the continuum-forming region. $r_c$ is the radius of the accretion column. Solid arrows denote the velocity of bulk motion with the velocity $\beta= v/c$, $c$ is the light speed. Wavy arrows denote diffusive flux emerging from the continuum-forming region. At $L= 2\times 10^{37}\,\mathrm{erg~s^{-1}}$, the bulk velocity in the line-forming region is decelerated significantly by radiation pressure due to resonant scattering.} 
    \label{Fig1}
\end{figure}

Assuming a radiation dominated accretion column, we conduct a Monte Carlo (MC) simulation to deal with the radiation transfer process inside the accretion column. 
We adopt a model derived by \cite{Arons1992ApJ} for an accretion mound shape as shown in Figure~\ref{Fig1}.
For $ 0 \leq r_{\perp}  < r_c$, the equality between the ram and radiation pressure yields the nominal location of the shock to be at the height 
\begin{equation}
h_s(r_{\perp})= R_{NS} \frac{L_a}{L_{ED}H_{\parallel}}(1-\frac{r_{\perp}^2 }{r_c^2}) \equiv h_{top}(1-\frac{r_{\perp}^2 }{r_c^2}), 
\label{Arons}
\end{equation}
where $r_{\perp}$ is the cylindrical radius with respect to the magnetic axis, 
$L_a=GM\dot{M}/R_{NS}$ is the accretion luminosity of one pole where $G$ is the gravitational constant,  $L_{ED}$ is Eddington luminosity.
$H_\parallel \sim H_\perp \sim1$ is a good approximation over most of the mound's volume \citep{Arons1992ApJ}. 
We consider a typical supercritical luminosity case with $L= 2\times 10^{37}\,\mathrm{erg~s^{-1}}$, implying a shock height of $\sim 1 \times 10^3$\,m \citep{Arons1992ApJ}. 
In the critical luminosity, the mound is expected to be as high as $r_c=10^3$\, m \citep{Basko1976MNRAS}.

The line-forming region is assumed to be a narrow region around an accretion mound with a height of $\sim 1 \times 10^3$\,m, thus we actually compute the cyclotron line in altitude from $\sim$ 10\,m to $\sim 1.1 \times 10^3$\,m in an accretion column with the surface cyclotron energy being assumed as $75\,$keV, such that the re-constructed phase-averaged spectra would produce two cyclotron lines at $\sim33\,$keV and $\sim55\,$keV respectively (coincide with observations, see \citealt{Ding2021MNRAS}). The strength of
the surface magnetic field can be about $6.5 \times 10^{12}\,$G when gravitational redshift is taken into account. 

The bulk velocity is expected to be decelerated  due to resonant scattering in the line-forming region as shown in Figure~\ref{Fig1}.
We assume the bulk velocity profile in the line-forming region is given by
\begin{equation}
v(z,r_{\perp}) = - \left[ v_{ff}(z_{max}) - \left\{ v_{ff}(z_{max})-v_0 \right\} \sqrt{\frac{z_{max}-z}{z_{max}-h_s(r_{\perp})}} \right],
\label{velocity}
\end{equation} 
where $z$ is the distance from the surface of a NS along the column axis, $v_{ff}(z)=-\sqrt{2GM_{NS}/(R_{NS}+z)}$ is free-fall velocity at a height $z$, $v(z)$ is decelerated from  $-v_{ff}(z_{max})$ to $-v_0$ in the line-forming region where $v_0=0.1c$ is the velocity at the surface of an accretion mound, making use of an approximate velocity profile $v \propto \tau \propto \sqrt{z}$ \citep{Becker2007ApJ}. Here, $c$ is the speed of light.

\begin{figure*}
	\centering
	\includegraphics[width=\textwidth]{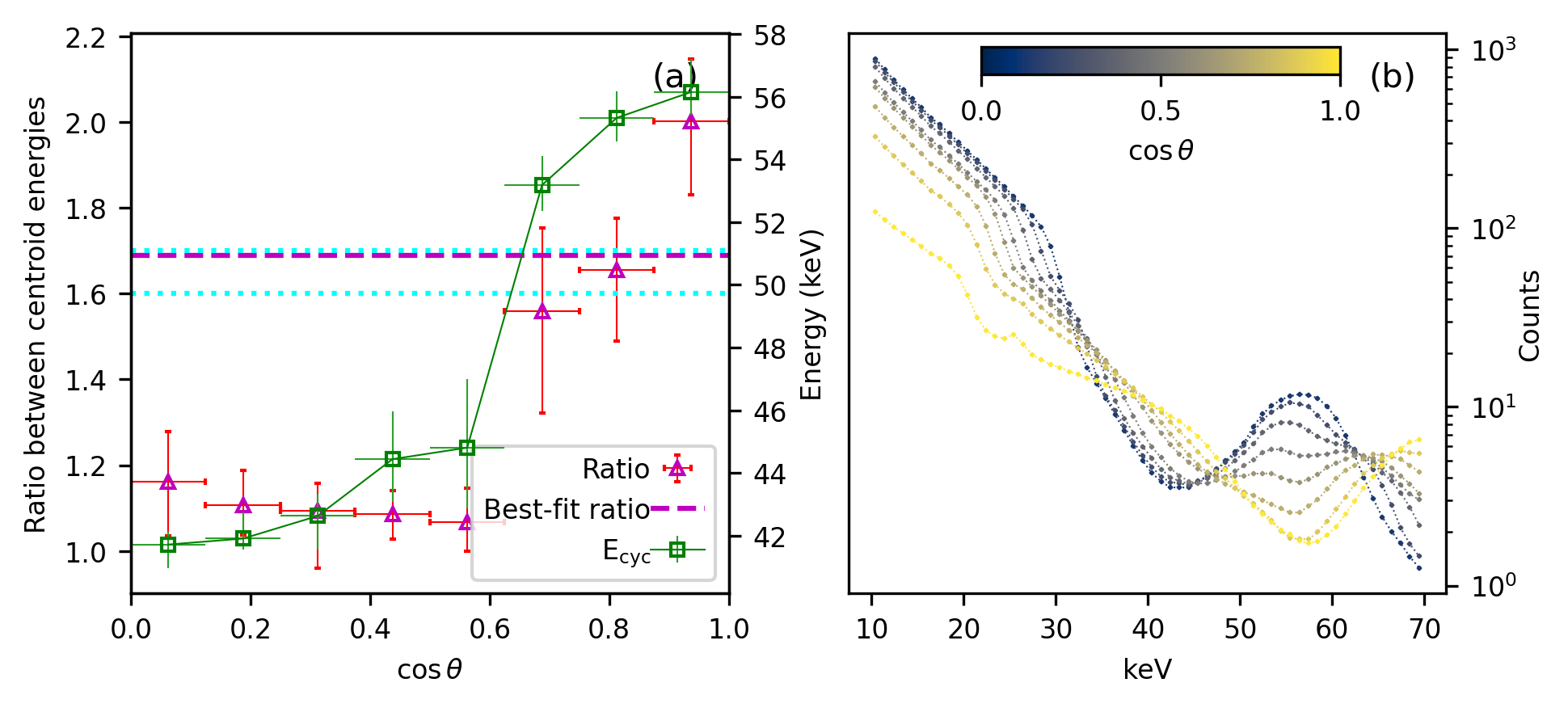}
	\caption{The simulated spectra and the centroid energy ratios of different photon emission angles. ({\bf a}) Centroid energy ratio between the two cyclotron absorption lines (y-axis), fitted from each simulated spectrum in a given photon emission angle ($\theta$, x-axis), which is the angle between the photon's motion direction and the magnetic field when the photon interacts with the line forming region. E$_\mathrm{cyc}$ is the centroid energy of the higher energy line in two observed cyclotron absorption lines. The dashed line is the best fitted ratio from the observations, which will be discussed later. ({\bf b}) Simulated spectra produced in different emission angles. The absorption core shifts with the local emission angle in $\sim 40-60\,$keV due to Doppler boosting and anisotropic beaming pattern.}
	\label{fig:simuspec}
\end{figure*}

We consider a beam pattern that have a beam with a peak intensity of $\theta_{p1}=60^{\circ}$, a pencil-like beam,  or a beam pattern composed of two beam patterns that have a beam with a peak intensity of $\theta_{p1}=60^{\circ}$ and with a peak intensity of $\theta_{p2}=90^{\circ}$, a fan beam.
Here we consider the angle-dependence of intensity of the continuum spectra, i.e., beam patterns, as follows \citep{Kraus2001ApJ}:
\begin{equation}
I(\theta)=\left(\exp\left[ -\frac{(\theta-\theta_{p1})^2}{d^2}\right]+\exp\left[ -\frac{(\theta-\theta_{p2})^2}{d^2}\right]\right)
\label{intensity}
\end{equation}
Here, $\theta$ is the angle between the direction of photon's propagation and B-field.
We use $60^{\circ}$ for parameter $\theta_{p1}$ and $90^{\circ}$ for  $\theta_{p2}$ respectively and fix $30^{\circ}$ for parameter $d$ as in previous theoretical papers \citep{nishimura2015},
since a pencil-like plus fan beam is considered 
in this work.
The radiation spectrum energy distribution (SED) is modeled by NPEX, which performs well both in the phase-averaged spectra \citep{Ding2021MNRAS} and in the phase-resolved spectra (see Section~\ref{sec: obs results}), being produced with two cut-off power-law components. As previous study has confirmed the lack of variability for the continuum parameter throughout the NS orbital motion, we set these parameters to be the mean values observed in previous phase-averaged study. Specifically, the low energy power-law component (Wien tail) is assumed to be a thousand time stronger than the hard component, while the photon index and high energy cut-off are frozen at 1.2 and $7.0\,$ keV respectively. Eight derived spectra, which obtained by uniformly distributing $\cos\theta$ from 0 to 1, are shown on the right panel of Figure~\ref{fig:simuspec}.

\subsection{Generation of the simulated phase-resolved spectra}

\begin{figure*}
	\centering
	\includegraphics[width=0.7\textwidth]{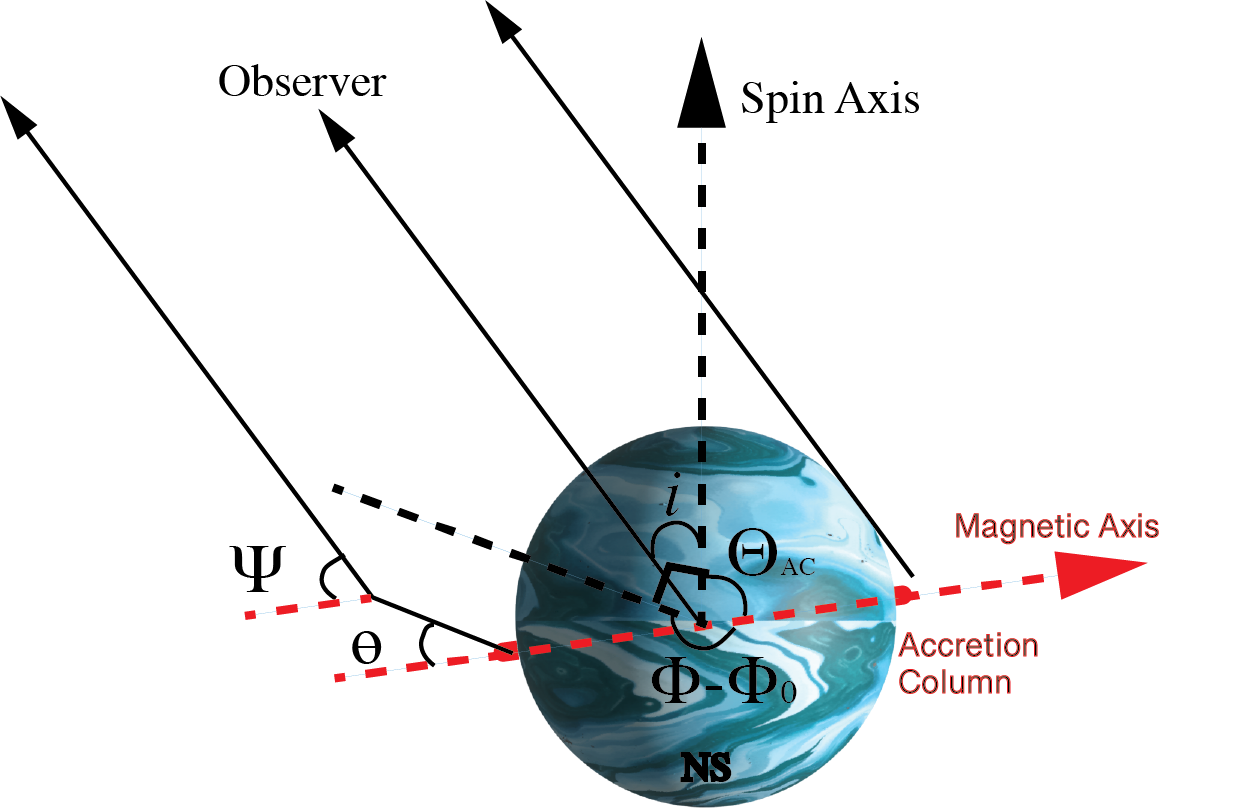}
	\caption{Main axes and angles of an accreting NS. The plane determined by the spin axis and observer is indicated with the black dashed line. The angles shown are $\Psi$ (apparent photon emission angle), $\Theta_{\rm AC}$ (magnetic inclination), $\theta$ (photon interacting angle with the line forming regions of accretion column), $i$ (NS inclination) and $\Phi-\Phi_{0}$ (angle between the planes characterising the initial and instantaneous spin phase).}
	\label{fig:system}
\end{figure*}

A photon emitted at an angle $\theta$ with respect to the radial direction would escape to infinity at a larger angle $\Psi$ due to general relativistic (GR) light bending near a compact object \citep{Beloborodov2002ApJ}:
\begin{equation}
	\cos\theta\approx1-(1-\cos\Psi)(1-R_{s}/R),
\end{equation}
\noindent
where $R_{s}$ is the Schwarzschild radius of the NS, while $R$ is the NS radius. We assume that the tidal stream from Wray 977, and the generated transient disk still enables significant alignment between NS spin and binary orbital angular momentum, thus the observer inclination with respect to the NS should be close to the binary inclination ($i\simeq66^{\circ}$, \citealt{Kaper2006A&A}), implying a geometric relation connecting the observed spin phases ($\Phi$) and the magnetic inclination of NS (see Figure~\ref{fig:system} for the definition of these angles):
\begin{equation}
	\cos \Psi=\cos i\cos \Theta_\mathrm{AC}+\sin i\sin\Theta_\mathrm{AC}\cos(\Phi-\Phi_{0}).
\end{equation}
\noindent
$\Theta_\mathrm{AC}$ here is the angle between accretion column and NS spin axis, being assumed to be identical to the magnetic inclination; $\Phi_{0}$ is an arbitrary phase offset. To deal with the problem of accretion columns being occluded by the NS, we assumed that as the accretion column crossing the limb of the star, disappearing from our line-of-sight, it quickly darkens and the polar cap at the opposite side will become dominant (i.e. we did not take the height of the column into account). Consequently, each combination of phase and $\Theta_\mathrm{AC}$ can be used to calculate the corresponding $\cos\theta$. Then, if a specific NS inclination is assumed, we could re-construct both the phase-averaged and phase-resolved spectra from the corresponding photon emission angle in any specific spin phase. Specifically, we assume the NS inclination is $66^{\circ}$. Same as the observational phase-resolved spectra, we set eight uniformly distributed phases. Additionally, we set $\Theta_\mathrm{AC}$ to be a arithmetic sequence ranging from 0 to 90 with an interval of 5. This allows us to generate a total of 8 * 19 spectra corresponding to different combinations of phases and $\Theta_\mathrm{AC}$. For each spectrum, we calculate the intensity by dividing its phase into eight equal width parts again (which gives a total of 64 uniformly distributed phases by now), and combining them with the $\Theta_\mathrm{AC}$ of the spectrum to calculate $\cos\theta$, determining which simulated spectrum (i.e. on the right panel of Figure~\ref{fig:simuspec}) each combination corresponds to. Subsequently, we average the eight simulated spectra to obtain the final simulated phase-resolved spectrum for each combination of phase and $\Theta_\mathrm{AC}$, with a total simulated phase-resolved spectra number of 8 * 19.

We then utilized the generated spectra by convolving them with authentic \textsl{Insight}-HXMT ancillary and response matrix, so as to show how these model spectra would appear when being observed with \textsl{Insight}-HXMT. To be more specific, we fit the simulated spectra with \textsc{NPEX}, while keep adding extra Gaussian absorption until no apparent structure is observable. After that, we loaded the ancillary and response file of \textsl{Insight}-HXMT and used the command \textsc{fakeit} built-in \textsc{Xspec} to produce fake spectra. The fake spectra are re-normalized so as to produce energy flux $\sim1\times10^{-8} \mathrm{erg~cm^{-2}~s^{-1}}$ in $10-70\,$keV, being comparable to radiation dominated state's value if consider the source distance. The generated HE spectra of 17 detectors are then combined using the same method provided in HXMTDAS v2.04. The resulting spectra are again fitted with NPEX models to derive the CRSF energy ratio. Such ratio, together with the centroid energy, help us compare the theoretical line profile with the observational one in the supercritical luminosity scenario. 
On the left panel of Figure~\ref{fig:simuspec}, we show the centroid energy ratio of the two CRSFs fitted from each simulated spectrum in a given photon emission angle. The result shows that the photon emitted at $\theta\lesssim 40^{\circ}$ ($\cos\theta> 0.65$, $\theta$ is the emission angle between photon and magnetic field) should be dominant in our observed spectrum.

\section{Results}

\subsection{Observational spectrum}
\label{sec: obs results}

\begin{figure*}
	\centering
	\includegraphics[width=0.7\textwidth]{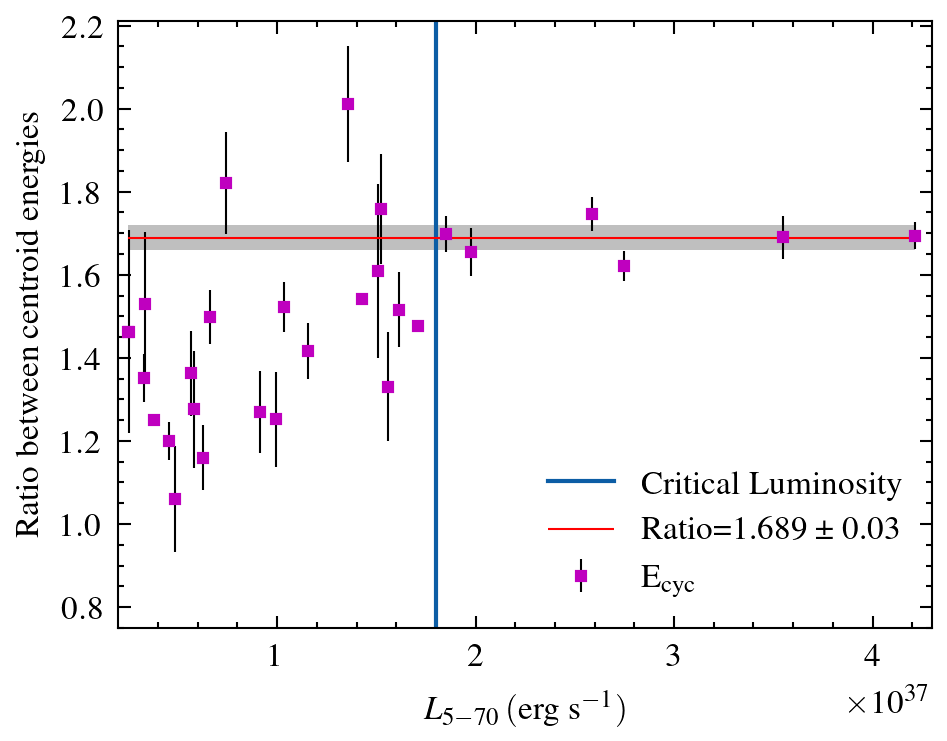}
	\caption{Centroid energy ratios in phase-resolved spectra between the two CRSFs of GX 301--2 based on all analyzed \textsl{Insight}-HXMT observations. The ratios are derived from phase-resolved spectra in each short pointing exposures, with typical net exposure time of $\sim3 $ks. Only ratios derived from more than four successfully fitted phase-resolved spectra in each observation are shown. The ratios converge to the value of $\sim 1.65$ implied in the results of phase averaged spectroscopy (also see Figure 13 in \citealt{Ding2021MNRAS} for details). The best-fit ratio, 1.69$\pm 0.03$, is derived from the super-critical state with higher luminosity ($\gtrsim 1.8\times 10^{37}\rm erg ~s^{-1}$ in $5-70$ keV, i.e. the vertical line).}
	\label{fig:ppratio}
\end{figure*}

As it is currently not practical to generate continuum from first principle in the context of accreting NS radiation simulation, we tested three different empirical continuum models first. \textsc{highEcut} is a high energy cut-off model, forcing the part above cut-off energy to decay exponentially:
\[
M_\mathrm{highEcut}=
\begin{cases} 
	\exp [(E_\mathrm{cut}-E)/E_\mathrm{fold}] &(E>E_\mathrm{cut}) \\
	0\ &(E\leq E_\mathrm{cut})
	\label{equ:highecut}
\end{cases} 
\]
However, due to the discontinuous derivative, it will produce an artificial cusp at $E_\mathrm{cut}$, which was modeled in this work with a Gaussian absorption. \textsc{newHcut} \citep{Burderi2000} uses polynomials to smooth the part around $E_\mathrm{cut}$, but introduced extra parameters like smooth range $\Delta E$, which is fixed to 5\,keV here. \textsc{NPEX} \citep{Mihara1995PhDT} is a combination of two cut-off power law models with common $E_\mathrm{cut}$, having a soft component (photon index fixed to 2) and a hard component, which generally outperforms other two models when being applied to averaged spectrum \citep{Ding2021MNRAS}. Besides these empirical continuum model, we also include a theoretical model for comparison, namely Becker-Wolff self-consistent cyclotron line model \citep[i.e. {\em bwcycl} in XSPEC][]{Becker2007ApJ,Ferrigno2009A&A}. Upon comparing the phase-resolved fitting results of these four models, we note that \textsc{NPEX} model outperforms the others, exhibiting more stable and normal parameters and yielding better chi-square values, especially when comparing with the derived parameters from \textsc{highEcut} and \textsc{newHcut}. On the other hand, {\em bwcycl} also performs good with its CRSF parameters very similar to those from the \textsc{NPEX} model, except for some phase-resolved spectra where the derived parameters of CRSFs and {\em bwcycl} differ significantly from those of other phase-resolved spectra in the same observation.  
Therefore, considering the quite stable results compared to others, NPEX has been chosen as our final continuum model. Besides, the final results of the centroid energy ratio of all the observations derived from the other three models are very similar the NPEX results, i.e., scattered below the critical luminosity and concentrated above it, with the best-fit ratios within the range of $\sim$ 1.6$-$1.8 as shown in Figure~\ref{fig:ppratio}.

\begin{figure*}
	\centering
	\includegraphics[width=1.0\textwidth]{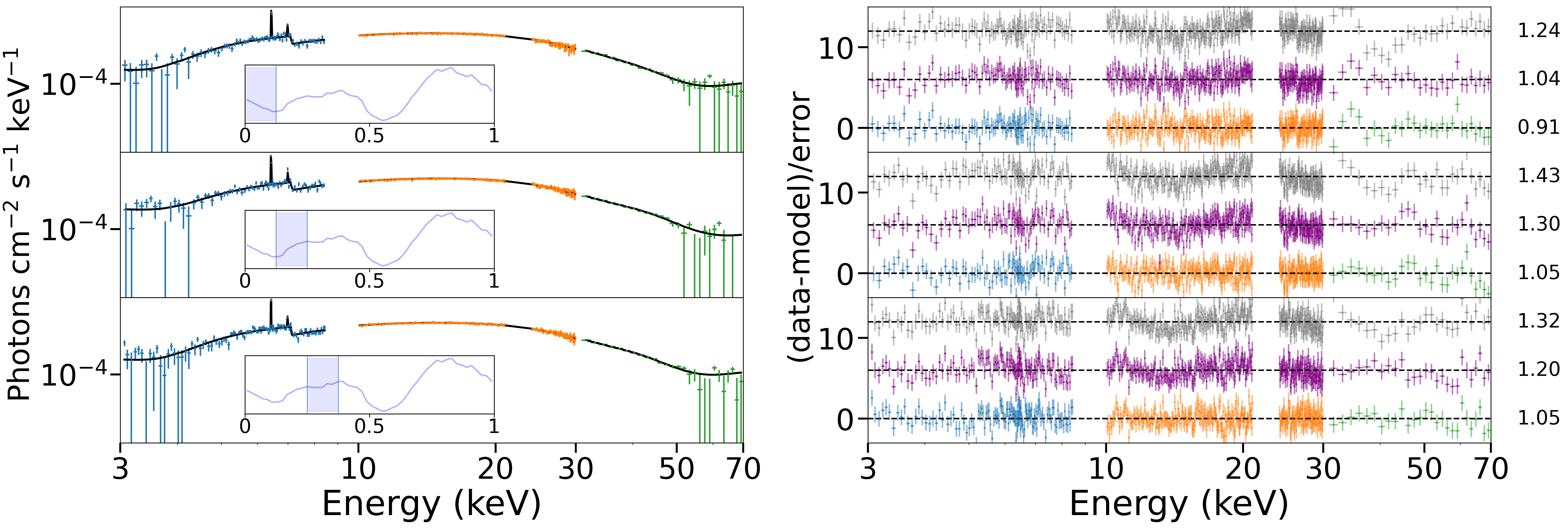}
	\caption{Phase-resolved spectra derived in a super-critical observation. The ObsID is P010130900103 with $L = 4.2 \times 10^{37}$ erg$/$s. Only results for the first three phases are presented, while the complete figure can be found in the appendix. {\bf Left panels:} Fitting spectra with pulse profiles inserted, highlighting the corresponding pulse phases in each panel. The model used here is {\em Constant*TBabs*(TBpcf*gabs*gabs*Continuum+Gaussian+Gaussian)}. {\bf Right panels:} Residuals corresponding to the left panels with the same colors. Residuals from model with only one {\em gabs} are shown in each panel with purple lines, and their values are all increased by 6, while residuals from model with no absorption are shown in each panel with grey lines, and their values are all increased by 12. The reduced chi-square value is shown on the right side for each model. The broad band spectral shape is stable throughout all pulse phases. Almost all complications are due to the cyclotron (hard band) and neutral stellar wind (soft band) absorption.}
	\label{fig:prspec}
\end{figure*}

\begin{figure*}
	\centering
	\includegraphics[width=1.0\textwidth]{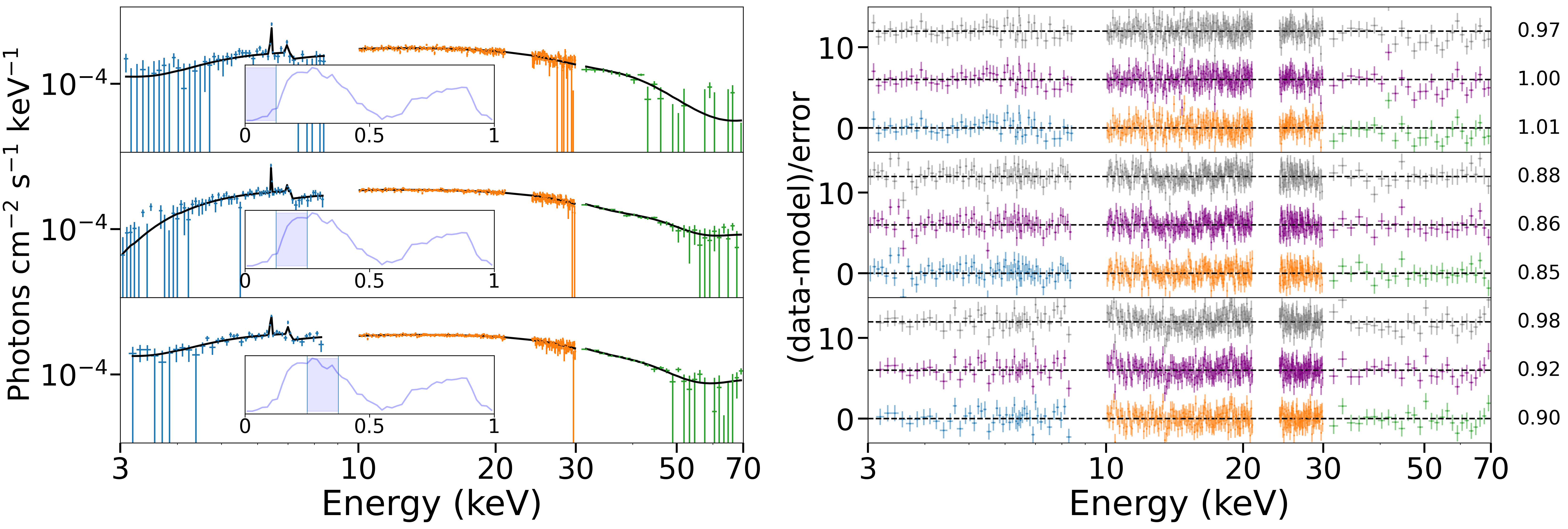}
	\caption{Phase-resolved spectra derived in a lower luminosity observation. The ObsID is P010130900401 with $L = 6.6 \times 10^{36}$ erg$/$s. Elements are the same as Figure~\ref{fig:prspec}. Only results for the first three phases are presented, while the complete figure can be found in the appendix.}
	\label{fig:prspec2}
\end{figure*}

The fitted spectral parameters of the phase-resolve spectral analysis for P010130900103 is presented in Figure~\ref{fig:prspec}, which is a super-critical case of $L = 4.2 \times 10^{37}$ erg/s. To confirm the existence of two CRSFs, residuals of models containing different number of {\em gabs} are shown on the right panel, with the corresponding reduced chi-square shown on the right side outside the panel, where the grey lines show the residuals from no gabs model, and the purple lines show the residuals from one gabs model. See the caption in Figure~\ref{fig:prspec} for a detailed description of the figure elements. Combining both the reduced chi-square and the residual structure, it is evident that the addition of each CRSF component noticeably improves the fitting results of the spectrum. A lower luminosity observation with $L = 6.6 \times 10^{36}$ erg/s is shown in Figure~\ref{fig:prspec2} for comparison. Although the fittings are still acceptable in lower luminosity cases, the results for several phases are not as good as in the super-critical observations. However, it is still evident that the inclusion of two CRSF models is necessary for the fitting in most phases, even though sometimes a second CRSF component only marginally improves the results.

\begin{figure*}
	\centering
	\includegraphics[width=1.0\textwidth]{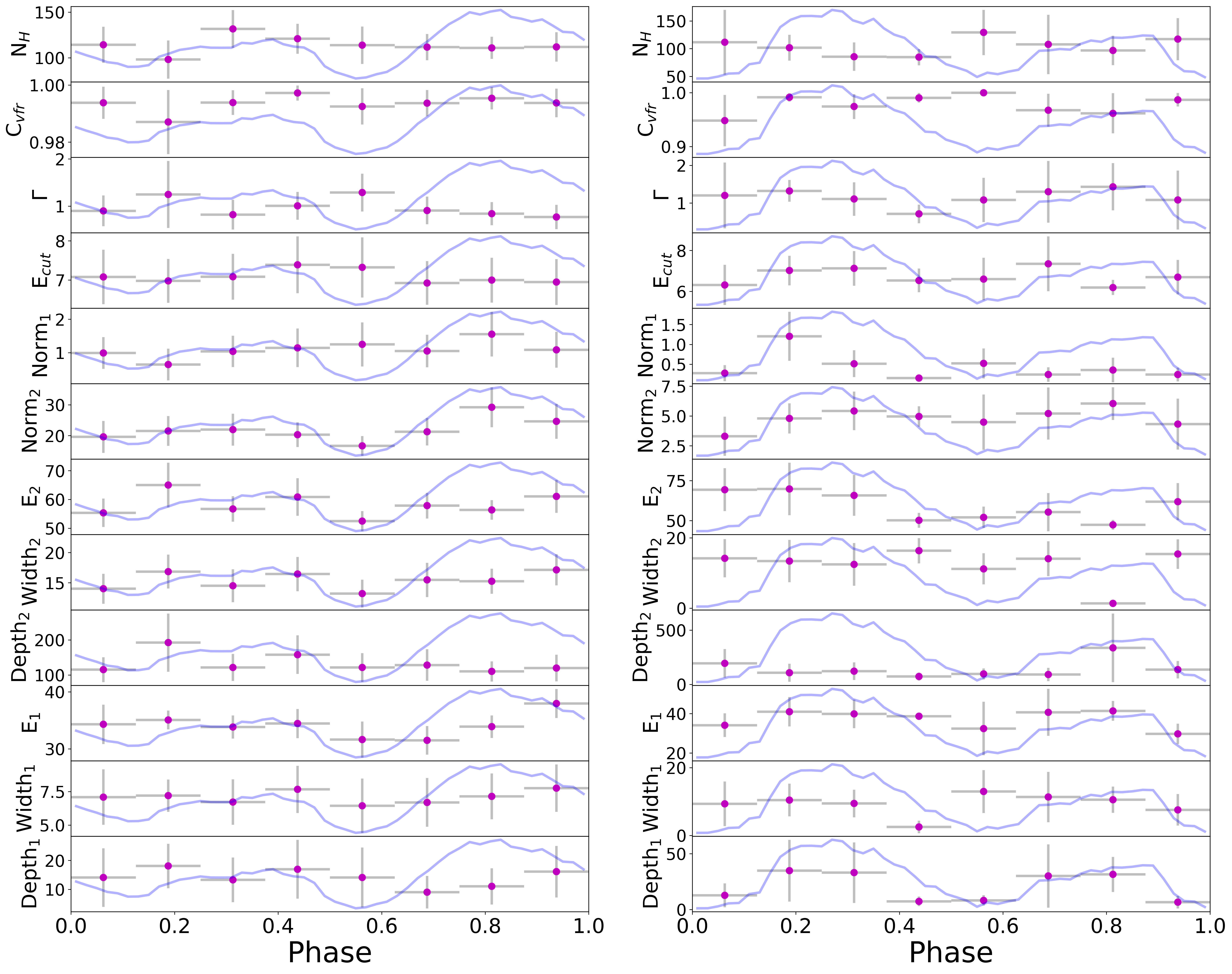}
	\caption{The evolution of the best-fit parameters with pulse phases. The parameter explanations and their units are the same as described in Table~\ref{tab:my-table}. ObsID P010130900103 results are shown on the left panels, which is a super-critical case, and obsID P010130900401 results with lower luminosity are shown on the right panels. Pulse profiles of the source in the HE band (30--60 keV) are shown in both sub panels with light blue lines.}
	\label{fig:pulse_conttinnum}
\end{figure*}

We present the evolution of the fitted parameters with their pulse phases in Figure~\ref{fig:pulse_conttinnum}. For the higher luminosity case of P01030900103, the energy of the second CRSF remains almost constant, with a slight decrease around the pulse minimum and an increase at the beginning of the second pulse, which resembles the result of the 2015 outburst observed by NuSTAR \citep{Furst2018A&A}, even though these variations are not significant considering the error bars. The CRSF at lower energy exhibits a subtle decreasing followed by increasing with phase, and the continuum parameter $\Gamma$ shows an inverse correlation with phase amplitude. All these results indicate a similarity in spectrum compared with the 2015 outburst, although the correlations presented by \cite{Furst2018A&A} is more pronounced. Meanwhile, the parameters of the lower luminosity observation showing on the right side of Figure~\ref{fig:pulse_conttinnum} exhibit different behaviors. The CRSF at higher energy decreases at the second pulse, and the CRSF at lower energy decreases when the source is around its pulse minimum. The photon index shows a positive correlation with phase amplitude. These different behaviors may suggest that spectrum below and above critical luminosity could have different properties, although they could also be attributed to the observational precision due to lower brightness. It is worth noting that the 2015 NuSTAR observation analyzed by \cite{Furst2018A&A} has a luminosity of around 2.8 $\times 10^{36}$ erg$/$s, which resembles our lower luminosity case. 
These fitting results are also shown in Table~\ref{tab:my-table} with their chi-square and degree of freedom presented. In addition, observation P010130900107 is also included in the table for comparison, which has a luminosity of $1.8 \times 10^{37}$ erg/s. The errors reported in this paper are generated with Markov Chain Monte Carlo (MCMC) method (50 walkers, 200000 steps), and are at 90\% confidence level except for special statements.

\begin{table*}
    \centering
    \footnotesize
    \caption{ Best-fit parameters for three observations in different spin phases, where P010130900103 has high luminosity of $4.2\times10^{37} erg/s$, P010130900107 has median luminosity of $1.8\times10^{37} erg/s$, and P010130900401 has low luminosity of $0.7\times10^{37} erg/s$.}
    \begin{threeparttable}
    \begin{tabular}{|p{0.5cm}|p{1.0cm}|p{1.4cm}|p{0.7cm}|p{0.7cm}|p{0.7cm}|p{0.7cm}|p{0.7cm}|p{0.7cm}|p{1cm}|p{1cm}|p{0.6cm}|p{0.8cm}|p{0.9cm}|}
        \hline
        (1) & (2) & (3) & (4) & (5) & (6) & (7) & (8) & (9) & (10) & (11) & (12) & (13) & (14)\\
        Phase & N$_H$         & $C_\mathrm{vfr}$        & $\Gamma$        & E$_\mathrm{cut}$        & Norm$_1$      & Norm$_2$           & E$_2$       & $Width_{2}$     & Depth$_{2}$     & E$_1$       & $Width_{1}$     & Depth$_{1}$ & chisq/dof    \\
        \hline
          & ($10^{22}$ cm$^{-2}$)&  &  &  (keV) &  & ($\times10^{-4}$) &  (keV) & (keV) &  & (keV) & (keV) & & \\ 
        \hline
        \hline
    \multicolumn{14}{|c|}{P010130900103} \\
0&114$\pm$20&0.994$\pm$0.006&0.9$\pm$0.3&7.1$\pm$0.7&1.0$\pm$0.5&20$\pm$5&55$\pm$5&14$\pm$2&115$\pm$35&34$\pm$3&7$\pm$2&14$\pm$10&356.8/393\\
0.125&98$\pm$21&0.987$\pm$0.011&1.2$\pm$0.7&7.0$\pm$0.6&0.6$\pm$0.5&22$\pm$5&65$\pm$8&17$\pm$3&192$\pm$83&35$\pm$2&7$\pm$1&18$\pm$8&408.7/390\\
0.25&132$\pm$21&0.994$\pm$0.004&0.8$\pm$0.3&7.1$\pm$0.6&1.0$\pm$0.5&22$\pm$5&57$\pm$4&15$\pm$3&122$\pm$39&34$\pm$2&7$\pm$2&13$\pm$8&450.6/428\\
0.375&121$\pm$16&0.997$\pm$0.003&1.0$\pm$0.3&7.4$\pm$0.7&1.1$\pm$0.6&20$\pm$4&61$\pm$7&16$\pm$3&158$\pm$55&34$\pm$3&8$\pm$2&17$\pm$10&446.4/435\\
0.5&114$\pm$20&0.993$\pm$0.006&1.3$\pm$0.4&7.3$\pm$0.8&1.2$\pm$0.7&17$\pm$3&52$\pm$3&13$\pm$2&121$\pm$41&32$\pm$3&6$\pm$2&14$\pm$10&374.1/415\\
0.625&112$\pm$14&0.994$\pm$0.005&0.9$\pm$0.3&6.9$\pm$0.6&1.0$\pm$0.5&21$\pm$4&58$\pm$4&15$\pm$3&128$\pm$45&32$\pm$3&7$\pm$2&9$\pm$6&467.0/461\\
0.75&111$\pm$12&0.995$\pm$0.004&0.8$\pm$0.2&7.0$\pm$0.6&1.6$\pm$0.7&29$\pm$6&56$\pm$3&15$\pm$2&110$\pm$29&34$\pm$2&7$\pm$2&11$\pm$6&451.8/435\\
0.875&112$\pm$16&0.994$\pm$0.005&0.8$\pm$0.3&6.9$\pm$0.6&1.1$\pm$0.5&25$\pm$6&61$\pm$6&17$\pm$3&120$\pm$38&38$\pm$3&8$\pm$2&16$\pm$9&417.9/411\\

\hline
\multicolumn{14}{|c|}{P010130900107} \\
0&112$\pm$18&0.997$\pm$0.003&1.0$\pm$0.4&6.4$\pm$0.6&0.6$\pm$0.4&8$\pm$2&69$\pm$11&16$\pm$4&213$\pm$111&37$\pm$6&9$\pm$4&14$\pm$12&368.8/420\\
0.125&110$\pm$13&0.996$\pm$0.004&0.8$\pm$0.2&6.5$\pm$0.6&0.6$\pm$0.2&8$\pm$2&61$\pm$8&14$\pm$5&146$\pm$71&32$\pm$3&5$\pm$3&8$\pm$6&424.6/432\\
0.25&98$\pm$11&0.995$\pm$0.005&0.7$\pm$0.2&7.1$\pm$0.5&0.5$\pm$0.2&11$\pm$2&54$\pm$2&11$\pm$2&73$\pm$21&34$\pm$1&6$\pm$1&11$\pm$5&393.1/431\\
0.375&101$\pm$13&0.995$\pm$0.005&0.7$\pm$0.2&7.0$\pm$0.4&0.6$\pm$0.3&12$\pm$2&63$\pm$8&17$\pm$3&134$\pm$50&37$\pm$2&6$\pm$1&14$\pm$7&446.9/447\\
0.5&112$\pm$5&0.997$\pm$0.003&0.8$\pm$0.1&6.7$\pm$0.3&0.7$\pm$0.1&8$\pm$1&70$\pm$11&18$\pm$2&80$\pm$33&41$\pm$2&9$\pm$2&24$\pm$12&398.7/427\\
0.625&122$\pm$27&0.993$\pm$0.007&1.0$\pm$0.4&6.6$\pm$0.6&0.6$\pm$0.3&8$\pm$2&72$\pm$11&15$\pm$4&234$\pm$127&37$\pm$4&8$\pm$3&23$\pm$16&371.7/415\\
0.75&104$\pm$18&0.994$\pm$0.006&0.7$\pm$0.3&6.9$\pm$0.5&0.4$\pm$0.2&9$\pm$2&54$\pm$4&11$\pm$4&81$\pm$31&35$\pm$3&6$\pm$2&13$\pm$10&426.9/422\\
0.875&117$\pm$23&0.992$\pm$0.007&0.7$\pm$0.3&6.8$\pm$0.7&0.3$\pm$0.2&11$\pm$4&55$\pm$5&15$\pm$4&91$\pm$44&31$\pm$3&6$\pm$6&3$\pm$2&355.5/403\\

\hline
\multicolumn{14}{|c|}{P010130900401} \\
0&112$\pm$58&0.948$\pm$0.048&1.2$\pm$0.9&6.3$\pm$1.0&0.3$\pm$0.2&3$\pm$2&69$\pm$13&14$\pm$5&194$\pm$131&34$\pm$6&9$\pm$7&13$\pm$11&309.7/308\\
0.125&102$\pm$23&0.992$\pm$0.008&1.3$\pm$0.3&7.0$\pm$0.7&1.2$\pm$0.6&5$\pm$1&70$\pm$16&13$\pm$6&108$\pm$83&41$\pm$7&10$\pm$5&35$\pm$28&296.3/349\\
0.25&86$\pm$26&0.974$\pm$0.023&1.1$\pm$0.4&7.1$\pm$0.8&0.5$\pm$0.3&5$\pm$2&66$\pm$13&12$\pm$6&122$\pm$82&40$\pm$7&9$\pm$4&33$\pm$27&302.0/337\\
0.375&85$\pm$15&0.991$\pm$0.009&0.7$\pm$0.2&6.5$\pm$0.6&0.2$\pm$0.1&5$\pm$1&50$\pm$5&16$\pm$4&73$\pm$37&39$\pm$1&3$\pm$2&7$\pm$4&239.8/347\\
0.5&129$\pm$41&1.000$\pm$0.000&1.1$\pm$0.6&6.6$\pm$1.0&0.5$\pm$0.4&4$\pm$2&52$\pm$7&11$\pm$4&96$\pm$50&32$\pm$14&13$\pm$6&8$\pm$5&301.4/346\\
0.625&108$\pm$53&0.968$\pm$0.030&1.3$\pm$0.8&7.4$\pm$1.3&0.2$\pm$0.2&5$\pm$2&55$\pm$12&14$\pm$5&91$\pm$62&41$\pm$12&11$\pm$7&30$\pm$28&336.0/335\\
0.75&97$\pm$27&0.962$\pm$0.037&1.4$\pm$0.6&6.2$\pm$0.4&0.4$\pm$0.3&6$\pm$1&47$\pm$3&1$\pm$1&338$\pm$316&41$\pm$5&11$\pm$4&32$\pm$16&343.6/353\\
0.875&117$\pm$38&0.987$\pm$0.013&1.1$\pm$0.8&6.7$\pm$0.8&0.2$\pm$0.2&4$\pm$2&62$\pm$12&15$\pm$4&136$\pm$80&30$\pm$5&8$\pm$5&7$\pm$6&299.0/346\\

        \hline
    \end{tabular}
\begin{tablenotes}
\footnotesize
\item[*] Col. (1): spin phase; Col. (2): local hydrogen number density; Col. (3) Covering factor; Col. (4) photon index; Col. (5) cut-off energy; Col. (6) cut-off power-law normalization in the soft component; Col. (7) cut-off power-law normalization in the hard component;  Col. (8) centroid energy of CRSF 2;  Col. (9) width of CRSF 2;  Col. (10) Depth of CRSF 2;  Col. (11) centroid energy of CRSF 1;  Col. (12) width of CRSF 1;  Col. (13) Depth of CRSF 1; Col. (14) chi square / dof
\end{tablenotes}
\end{threeparttable}
\label{tab:my-table}
\end{table*}

\begin{figure*}
    \centering
    \includegraphics[width=0.45\textwidth]{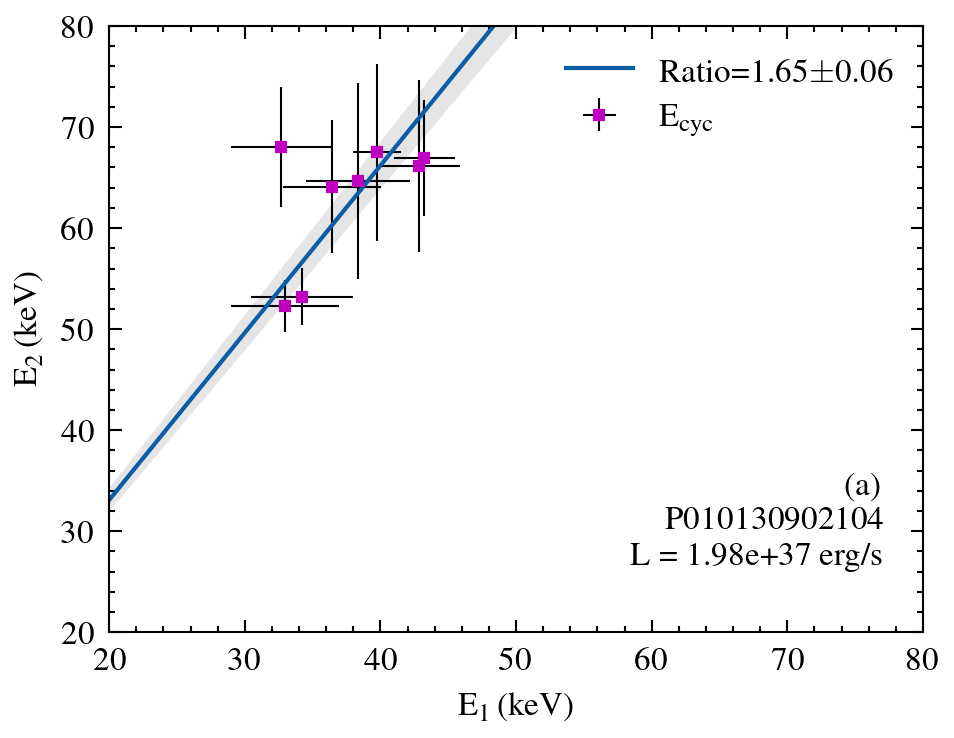}
    \includegraphics[width=0.45\textwidth]{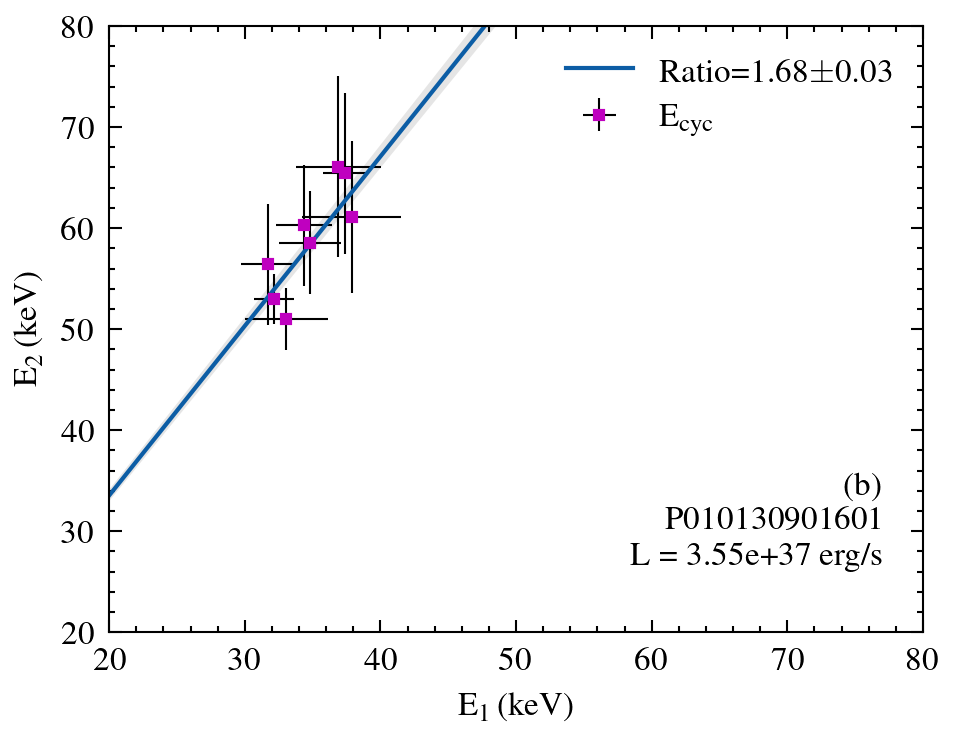}
    \includegraphics[width=0.45\textwidth]{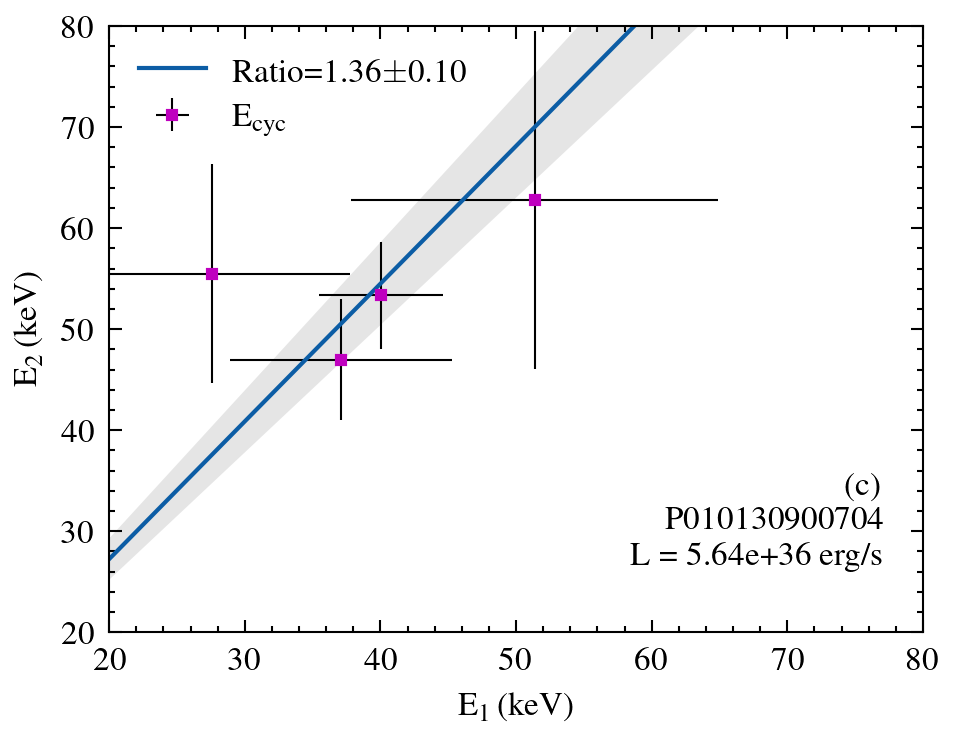}
    \includegraphics[width=0.45\textwidth]{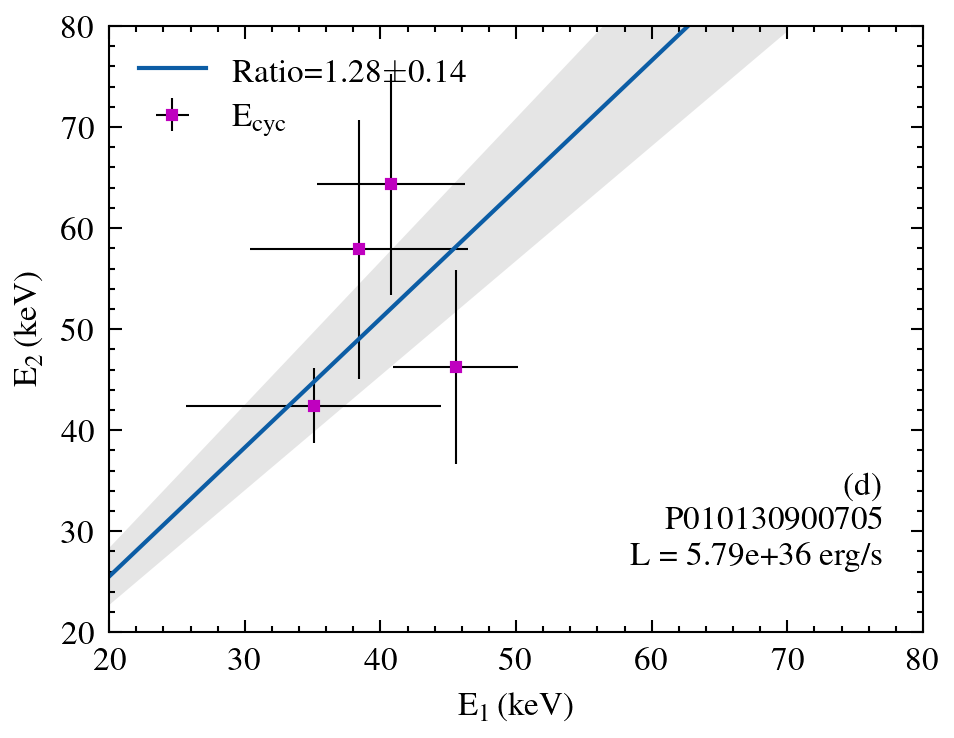}
    \caption{Cyclotron line centroid energies of the pulse-phase resolved spectral analysis derived from four pointing observation conducted from 2017 -- 2019. The ObsIDs are ({\bf a}) P010130902104; ({\bf b}) P010130901601; ({\bf c}) P010130900704; ({\bf d}) P010130900705. The fitted cyclotron line centroid energy ratio ($E_2/E_1$) and luminosity are shown on each sub-panel.}
    \label{fig:pr_ratio_sample}
\end{figure*}

To avoid the influence of inaccurate fitting on the results, we do not consider phases with fitted absorption line energies greater than 70 keV, which is beyond the energy range of our spectra. After that, the best-fit ratio ($E_2/E_1$) of two cyclotron line centroid energies is derived by Orthogonal Distance Regression (ODR) so as to consider errors in both variables. The samples are shown in Figure~\ref{fig:pr_ratio_sample}, where the upper two panels are for the super-critical cases and the lower two panels are for the low luminosity cases. We can see that the upper ones are more concentrated with smaller errorbars, while the lower ones have large error bars and are scattered.

\subsection{Simulated spectrum}

\begin{figure*}
	\centering
	\includegraphics[width=0.6\textwidth]{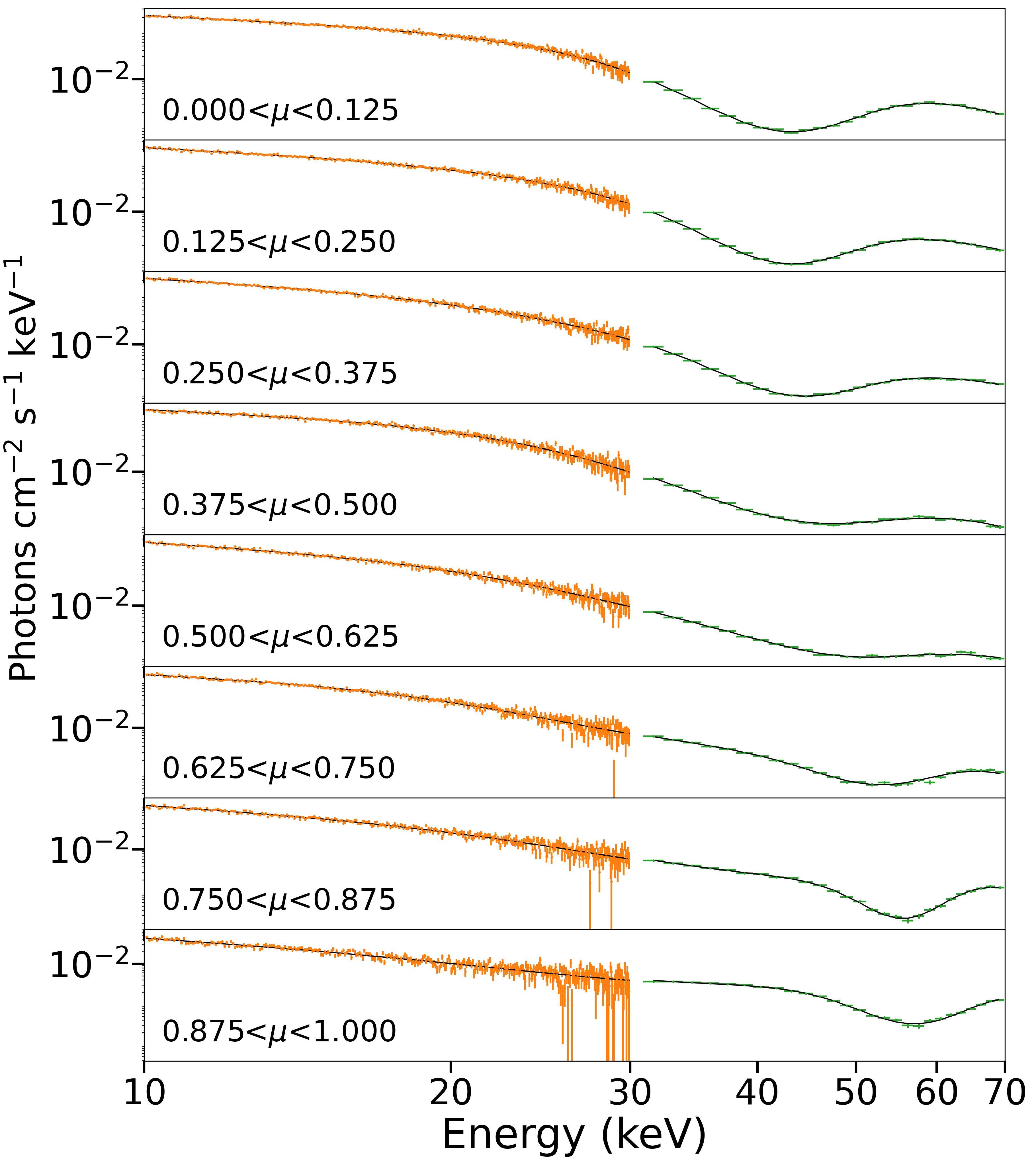}
	\caption{MC simulated spectra convolved with \textsl{Insight}-HXMT response matrix and background viewing from different angles. We assume the exposure is 150$\,$s, and the gravitational light bending effect is not considered. We generated spectra in $10-70\,$keV range and fitted them with the same observational model without neutral absorption from interstellar medium and stellar wind, where $\mu$ in the sub panels is equal to $\cos\theta$.}
	\label{fig:simuspecPR}
\end{figure*}

It has been argued that the complex CRSFs in GX 301--2 may not result from the presence of two separate line forming regions, but entail multiple line forming regions \citep{Ding2021MNRAS}. Specifically, recent theoretical models could naturally produce such absorption profiles \citep{nishimura2015,Schwarm2017A&A}. Using NPEX model as the input spectra and a fan-like input beaming pattern, we can successfully reproduce the observed structures, which are presented on the right panel of Figure~\ref{fig:simuspec} with different emission angles. To illustrate how they appear when observed by \textsl{Insight}-HXMT, they are also convolved with authentic \textsl{Insight}-HXMT ancillary and response matrix. They are then fitted with NPEX model plus two CRSF components, where the NPEX parameters are fixed to the values used to generate the spectra as discussed in Section~\ref{sec:simu-specModel}. Figure~\ref{fig:simuspecPR} and Tabel~\ref{tab:simuData} show the fitted results.

\begin{table*}
	\centering
	\caption{ Best-fit parameters with uncertainties for the MC simulated spectra as shown in Figure~\ref{fig:simuspecPR}. The {\em NPEX} parameters are not shown in the table because they are fixed to the values used to generate the simulated spectra, i.e. the photon index and high energy cut-off are 1.2 and 7.0 keV respectively.}
	\label{tab:simuData}
	\begin{tabular}{ccccccccc}
		\hline
		\hline
$\cos\theta$&Ecyc1&Width1&Depth1&Ecyc2&Width2&Depth2&norm&chisq/dof\\
0&41.8$\pm$0.2&7.6$\pm$0.2&49.2$\pm$2.3&35.6$\pm$0.8&18.6$\pm$1.3&72.0$\pm$9.9&0.010$\pm$0.001&321.0/359\\
0.125&42.0$\pm$0.3&6.5$\pm$0.3&33.4$\pm$2.4&37.8$\pm$0.5&17.8$\pm$1.7&88.1$\pm$12.3&0.010$\pm$0.002&356.3/359\\
0.25&38.8$\pm$0.4&19.2$\pm$0.7&128.2$\pm$8.3&42.8$\pm$0.3&6.6$\pm$0.4&24.5$\pm$2.1&0.014$\pm$0.002&333.0/359\\
0.375&44.2$\pm$2.1&7.8$\pm$3.7&32.3$\pm$27.8&37.9$\pm$3.2&15.8$\pm$3.4&54.8$\pm$25.2&0.005$\pm$0.001&366.4/359\\
0.5&38.6$\pm$2.2&18.6$\pm$1.2&125.4$\pm$21.2&48.5$\pm$2.3&9.6$\pm$4.0&23.6$\pm$18.8&0.010$\pm$0.001&364.3/359\\
0.625&35.9$\pm$2.2&15.4$\pm$2.1&104.9$\pm$25.1&53.8$\pm$0.7&6.8$\pm$1.1&27.6$\pm$10.1&0.008$\pm$0.002&386.4/359\\
0.75&55.4$\pm$0.4&5.4$\pm$0.8&28.4$\pm$3.6&34.9$\pm$0.9&18.7$\pm$1.3&145.6$\pm$18.9&0.011$\pm$0.002&390.7/359\\
0.875&56.4$\pm$0.7&7.2$\pm$0.8&44.6$\pm$7.3&28.0$\pm$1.0&18.9$\pm$1.0&172.8$\pm$21.9&0.013$\pm$0.004&372.0/359\\
\hline
\end{tabular}
\end{table*}

\begin{figure*}
    \centering
    \includegraphics[width=0.9\textwidth]{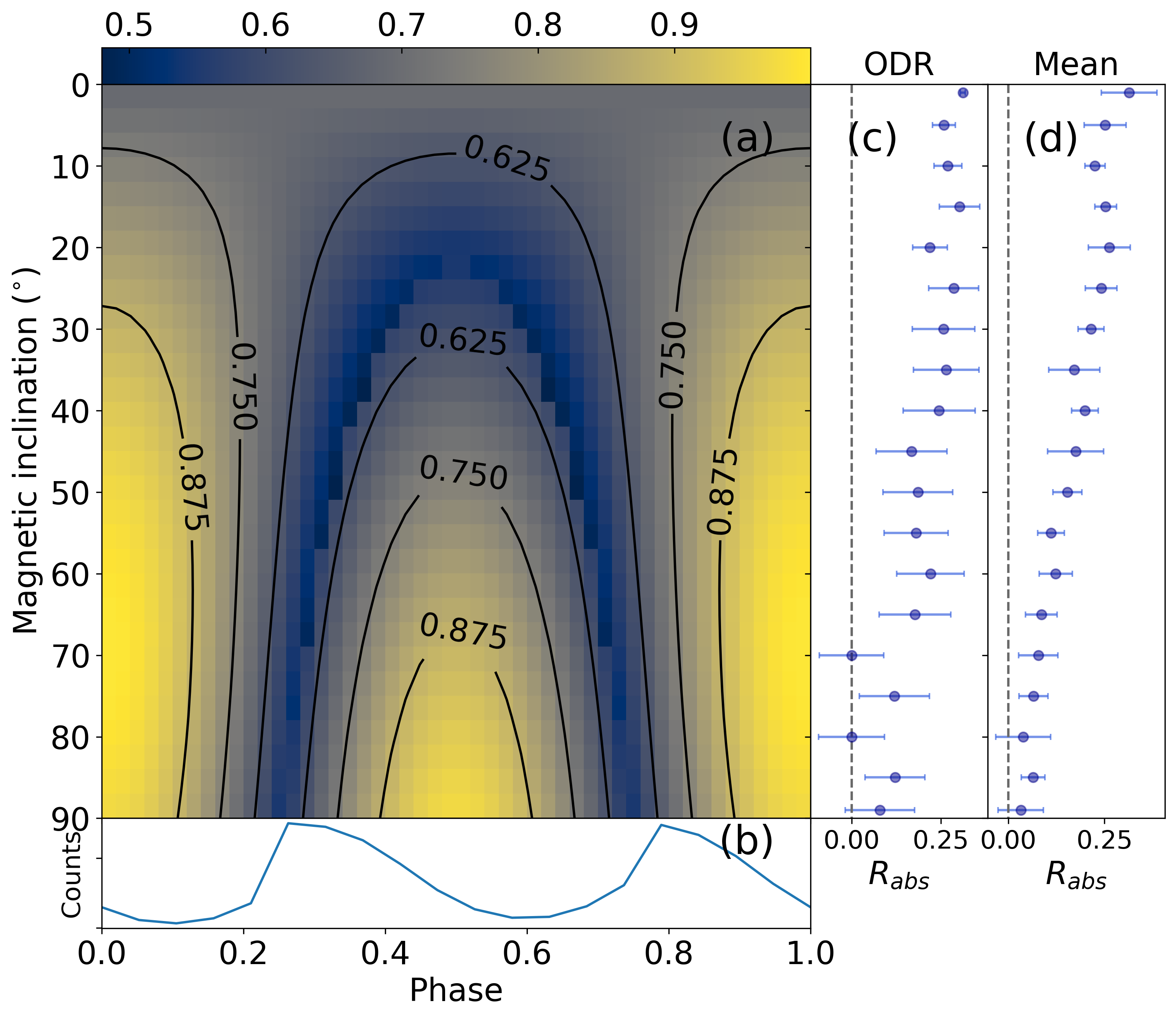}
    \caption{Fitting results to simulated spectra after combining with \textsl{Insight}-HXMT ancillary and response matrix. ({\bf a}) A heat map shows the photon interacting angle ($\cos\theta$) with the line forming regions. We could observe these photons with a given magnetic axis (y-axis) in different spin phases (x-axis). The simulated spectra in Figure~\ref{fig:simuspec}(b) are integrated into each phase bin in this map. After assuming a specific magnetic inclination, we are able to use the simulated spectra to re-construct phase-averaged and phase-resolved spectra. ({\bf b}) The generated input pulse profile when magnetic inclination is $85^{\circ}$. ({\bf c}) Fitting results (ODR) of simulated phase-resolved spectra with different magnetic inclinations, where $R_{abs}$ is the difference between the ratio of two cyclotron lines from the simulated spectra and the observed ratio obtained in Figure~\ref{fig:ppratio}.  ({\bf d}) Similar with (c) except the ratio from the simulated spectra is calculated using the average instead of ODR. All errors indicated in this figure are in 68\% confidence level (1$\sigma$). Spectral simulations imply that the NS is a nearly orthogonal rotator with a magnetic inclination $\Theta_{\rm AC} \gtrsim 70^{\circ}$.} 
    \label{fig:MagInc}
\end{figure*}

The resulted ratio in phase-averaged spectra is in good agreement with observation ($\sim 1.6$, but has no apparent variation with magnetic inclination), while the re-constructed phase-resolved spectra show strong dependency on magnetic inclination. In Figure~\ref{fig:MagInc}, we show the fitting results of the simulated spectra, where $R_{abs}$ is the difference between the ratio of two cyclotron lines from the simulated spectra and the observed ratio obtained in Figure~\ref{fig:ppratio}. The result shows that the observed photon emission angle is seriously restricted due to the magnetic geometry of the NS and the GR effect. We found that to obtain the CRSF structure in the phase-resolved spectral observations, the NS magnetic inclination should take special values. Assuming different magnetic inclination angles, the phase-resolved spectral simulations and fittings are performed to derive the cyclotron line centroid energy ratio. The results imply that the magnetic inclination of the NS should have the solution of $\gtrsim 70^{\circ}$, i.e. a nearly orthogonal rotator, with $68\%$ confidence level. The centroid energy ratio between the two CRSFs in GX 301--2 is the result of a near-orthogonal magnetic pole relative to the spin axis of the NS.

\section{Discussions}
\label{sec2:result}

\subsection{Phase-resolved spectra from \textsl{Insight}-HXMT}
With frequent \textsl{Insight}-HXMT observations, we performed phase-resolved spectral analysis of GX 301--2, confirming the existence of two CRSFs in all the phases independent of the continuum spectrum models. The centroid energy ratio ($E_{2}/E_{1}$) between the two CRSFs with the X-ray luminosity is plotted (Figure~\ref{fig:ppratio}), discovering the ratio $E_{2}/E_{1}\sim 1.6-1.7$ in the super-critical state ($L\gtrsim 1.8\times10^{37}\rm \,erg~s^{-1}$). The absorption structures in the phase-resolved spectra are modeled by Gaussian absorption components while the continuum is described with the empirical models, i.e., the Negative Positive Exponential \citep[NPEX,][]{Mihara1995PhDT} model. 
The centroid energy ratio has a large scatter from $1.2-2$ in the low luminosity, while reaches a narrow value range of $\sim 1.6-1.7$ in high luminosity, and this value is well consistent with the observed ratio in the averaged spectra \citep{Ding2021MNRAS}. This is remarkable because it means that the fixed energy ratio observed previously in phase averaged spectra is not due to averaging between different absorption profile, but is the consequence of the appearance of similar spectral structures in different spin phases. Such dependency appears most prominently in high luminosity state, with the critical luminosity possibly being the turning point ($\sim 1.8\times10^{37}\,\mathrm{erg~s^{-1}}$, \citealt{Ding2021MNRAS}). It is currently not clear whether this is predominantly due to the more stable radiation dominated accretion column structure above the critical luminosity, because \textsl{Insight}-HXMT is not capable of fully resolving the CRSF structure in the phase-resolved spectra at low luminosity ($\lesssim 10^{37}\rm \,erg~s^{-1}$). For observations with energy ratio close to unity where it is not easy to separate one line from the other, the uncertainties of the fitted absorption energies at low luminosity are large compared to those above the critical luminosity, resulting in larger errorbars in the energy ratio as shown in Figure~\ref{fig:ppratio}. Thus, the scattering and decrease of energy ratio could be the consequence of degeneration between two Gaussian absorption components in low signal-to-noise ratio.

We may note that the first CRSF is in the energy range of $\sim 30 - \sim 45$ keV considering the uncertainties as shown in Table~\ref{tab:my-table}. For a first CRSF at 45 keV, the upper limit of 70 keV chosen in our spectra analysis can cover only an energy ratio of 1.5, which may seriously affect the final results. However, the HE detector of \textit{Insight}-HXMT can reach 250 keV at its maximum \citep[e.g.][]{Ma2021}, and has been used for CRSF research up to 100 keV \citep{Liu2022MNRAS.514.2805L}. Consequently, the spectra fitting of 3-100 keV has been tested. The energy ratio remains stable compared to the 3-70 keV fitting results for high luminosity observations, and fluctuated for low luminosity observations. Most importantly, the centroid energy of the second CRSF do not exceed 70 keV even for the low luminosity cases (not include the uncertainties), and the energy ratios of two CRSFs for all observations exhibit similar characteristics: scatter in the low luminosity, but concentrate in the high luminosity, with the best-fit ratio of the super-critical state derived to be 1.64$\pm$0.02. However, the parameters derived from 3-100 keV is unstable with larger uncertainties, and the spectra above 70 keV is noisy, following the averaged spectra analysis performed in \cite{Ding2021MNRAS}, we keep the 3-70 keV results.

\begin{figure*}
	\centering
	\includegraphics[width=0.7\textwidth]{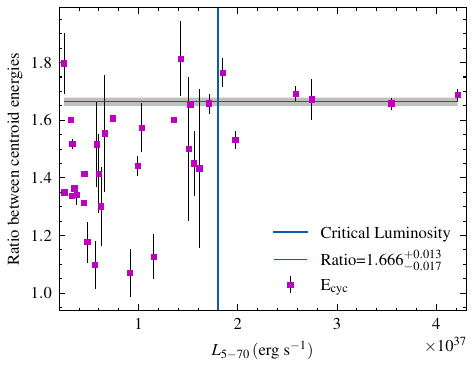}
	\caption{Centroid energy ratios in phase-resolved spectra between the two CRSFs of GX 301--2 based on all analyzed \textsl{Insight}-HXMT observations. The results are derived with four pulse phases of 0.15-0.4, 0.4-0.6, 0.6-0.9, 0.9-1.15. Elements can be referred in Figure~\ref{fig:ppratio}.}
	\label{fig:ppratio_4phase}
\end{figure*}

To check if the results are affected by different separated phases, we recalculated the centroid energy ratios in phase-resolved spectra with different pulse phases, i.e., 0.15-0.4, 0.4-0.6, 0.6-0.9, 0.9-1.15, which separate the pulse phases roughly based on the first peak, pulse minima and second peak. The results are shown in Figure~\ref{fig:ppratio_4phase}. Despite minor differences existed as compared with Figure~\ref{fig:ppratio}, the plot is still consistent with it, i.e., scattered below the critical luminosity and concentrated above it, and the best-fit ratio is around 1.67, which is within the error bar of our 8-phases results. We need to point out that only ratios with more than four successfully fitted phase-resolved spectra in each observation are shown in Figure~\ref{fig:ppratio} for more reliable results, and ratios fitted from less phases as shown in Figure~\ref{fig:ppratio_4phase} may raise inaccuracy.

\begin{figure*}
	\centering
	\includegraphics[width=0.45\textwidth]{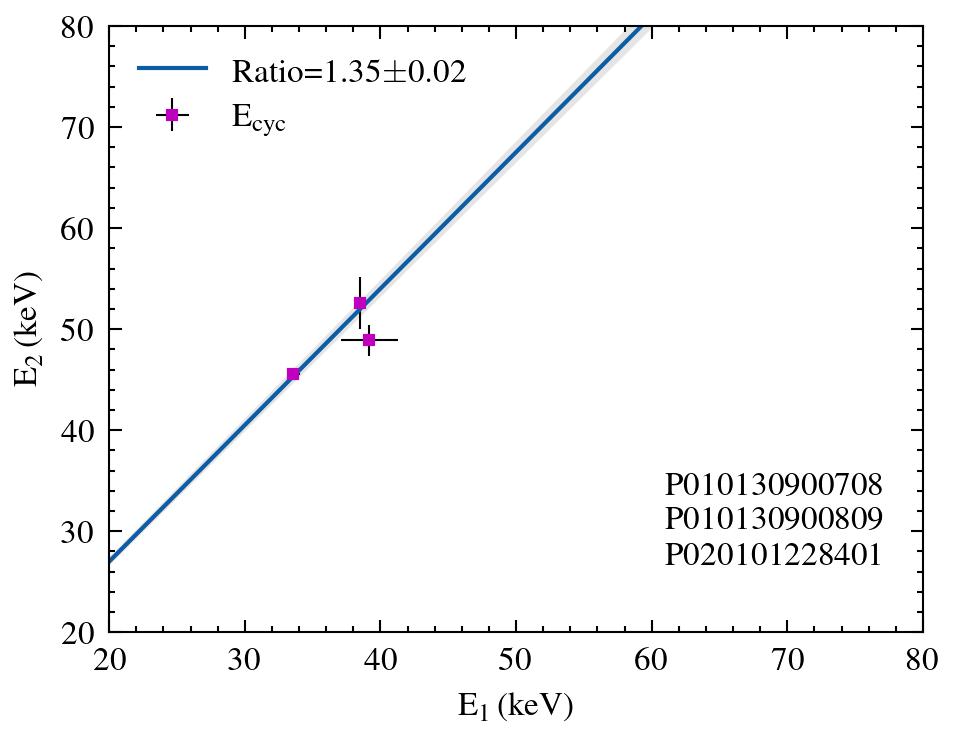}
	\includegraphics[width=0.45\textwidth]{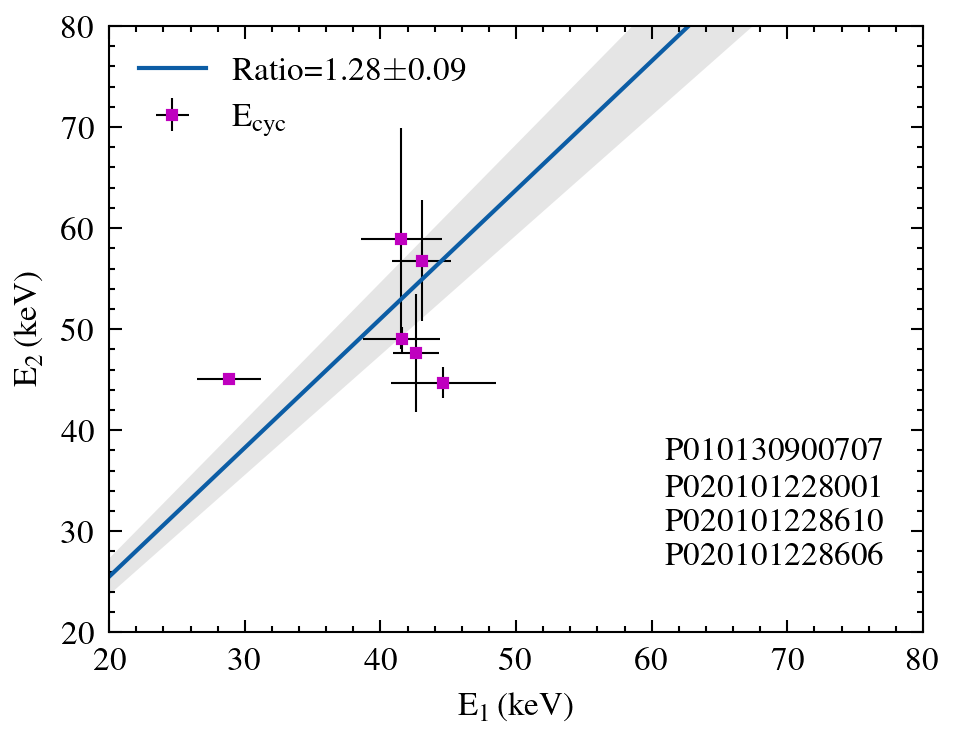}
	\caption{Cyclotron line centroid energies of the pulse-phase resolved spectral analysis derived from two combined spectra. The combined observations and fitted cyclotron line centroid energy ratio ($E_2/E_1$) are shown on each panel.}
	\label{fig:combine}
\end{figure*}

Combining multiple spectra may increase the confidence level of the results, thus we select observations with similar luminosity, and combine their phase-resolved spectra for each of the eight phases. The final results are shown in Figure~\ref{fig:combine}. On the left panel, we combine three observations of luminosity $\sim0.35\times10^{37}$ erg/s, and the right panel contains the combined results of four observations with their luminosity $\sim1.0\times10^{37}$ erg/s. Before combining the spectra, the energy uncertainties for different phases in these selected observations were quite large, and the energy distribution was scattered. Due to the low luminosity, some of these observations only had one or two successfully fitted phases, resulting in randomly distributed ratios ranging from $\sim$ 1.1 to $\sim$ 1.5. However, as shown in Figure~\ref{fig:combine}, the fitting results are significantly improved with much smaller uncertainties after combination, and the data points become more concentrated. The fitted ratios are both around 1.3, much lower than the best-fit ratio of the super-critical state. The enhanced spectra results again demonstrate the significant difference in the energy ratio above and below the critical luminosity.

\subsection{Magnetic inclination angle}

The CRSF is highly anisotropic due to the uneven velocity distribution and magnetic field strength inside the accretion column \citep{Becker1998ApJ,nishimura2015}. Consequently, the cyclotron absorption lines can provide important information about the material distribution near the NS magnetic polar region. For GX 301--2, two reported cyclotron absorption lines $E_1\sim 30-42$ keV and $E_2\sim 48-56$ keV show the similar relationships of the centroid energy versus X-ray luminosity \citep{Ding2021MNRAS}, suggesting a state transition at a critical luminosity of $\sim 1.8\times 10^{37}$ erg\ s$^{-1}$. To further investigate the properties of the two CRSFs and their relation to the magnetized NS, we model the accretion structure to study the simulated phase-resolved spectra. Based on the aforementioned observational results, we generate and fit the simulated spectra to study the physical origin of the CRSFs. 


Our findings provide strong constraints on the magnetic inclination angle in an accreting neutron star. Previously, by assuming a circular beam along the axis of the dipole magnetic field, the magnetic inclination angles of radio pulsars could be measured with polarization data \citep{Tauris1998MNRAS}. Without regard to the accretion torque, the evolution of magnetic inclination would be dominated by pulsar radiation spin down and inner viscous dissipation. Recent theoretical results indicate that isolated NS naturally evolves into two classes: near-aligned and near-orthogonal rotators - with typical pulsars falling into the latter category \citep{Lander2018MNRAS,Novoselov2020MNRAS}. This scenario was supported by the increasing separation between the main pulse and interpulse in Crab pulsar, implying the growth of the magnetic inclination angle \citep{Lyne2013Sci}.

However, the magnetic inclination angle in accreting X-ray pulsars is still poorly understood both in observations and theories. Preliminary studies deduced the magnetic inclination distribution from pulse profiles by assuming black-body radiation from a hot spot on the polar cap region \citep{Annala2010A&A,Leahy1990ApJ}. For accreting NSs in binary systems, the torque injection through either transient or prolonged accretion disc further complicates the physical picture. Magnetic angle evolution study in accreting NSs showed that the magnetosphere-disc interaction could orthogonalize the magnetic pole, and the magnetic inclination is greater than $55^\circ$ with all reasonable parameter combination \citep{Biryukov2021MNRAS}. By predicting a termination of inclination evolution after the disc-spin alignment stage, the calculation implies that the NS magnetic inclination evolution would be again dominated by inner (possibly bulk) viscosity once the disc is aligned with spin, which is likely to drive the NS into an orthogonal rotator \citep{Lander2018MNRAS,Biryukov2021MNRAS}. Our observations confirm that a slowly rotating and magnetized accreting NS in a binary system could indeed take an orthogonal configuration.

The radiation spectral properties (including the phase-resolved CRSF) of the NS accretion system GX 301--2 can be described by a radiation dominated accretion column of a $\sim 500\,$m shock height on the magnetic poles with a surface magnetic field strength of $B\simeq 8\times10^{12}\,$G. The NS magnetic field generally can be derived by $B=\frac{E_{\rm cyc}(1+z)}{11.57\ \mathrm{keV}} 10^{12} \mathrm{G}$, where $E_{\rm cyc}$ is cyclotron line energy, $z\sim 1.2$ is the gravitational redshift near the NS surface \citep{Staubert2019A&A}, which determines $B\sim (4-5)\times 10^{12}$ G for GX 301--2. This simple calculation should be an under-estimate because apart from the general relativistic redshift effect, the bulk motion in accretion column and real line production regions could significantly lower the resonant energy from the value produced by the corresponding local magnetic field \citep{nishimura2022}. Moreover, the width of CRSF is mainly determined by the bulk velocity gradient in our simulation and the depth of the absorption profile is influenced by beaming pattern, which can explain the wide absorption feature in GX 301--2 and other accreting X-ray pulsars \citep{Staubert2019A&A}. On the other hand, a less collimated (e.g. fan-like pattern) beaming pattern along with a smaller gradient would result in a broader and shallower line profile, providing a possible explanation for the non-detection of CRSFs in many accreting X-ray pulsars.


These rotating NSs in accreting systems could be continuous gravitational wave (GW) sources \citep{Bildsten1998ApJ}. It has already been shown that for an axially symmetric rotating fluid neutron star to emit GW, the rotation axis and the magnetic field axis should not be aligned to each other \citep{Bonazzola1996A&A}. An orthogonal rotator, due to its large toroidal $B$ field, is likely to have significant quadrupolar distortion, thus would act as the optimal configuration for GW emission \citep{Cutler2002PhysRevD}. The characteristic frequency of GW emissions from GX 301--2 would be around 3 mHz, which has the possibility of detection using future space-based GW telescopes like LISA (detailed calculations would be beyond the scope of this paper, would be resolved in next work). GW signals could provide insights into the magnetic inclination angle of GX 301--2 and other rotating neutron stars in accreting systems.

\section{Conclusions}

As a summary, phase-resolved spectral analysis is performed on the observation data of GX 301--2. Two CRSFs are detected in all phases, with their centroid energy ratio $\sim 1.6-1.7$ in the super-critical case, while it scattered in the range of $1.2-2$ when the source luminosity is lower. Based on the observed results, we simulate the phase-resolved X-ray spectra to determine the NS magnetic inclination. Our simulation shows that in order to obtain results similar to the observed data, the inclination angle needs to be greater than 70 degrees. This new method can be applied to other NS X-ray binaries and enable us to possibly derive the magnetic inclination angle distribution in different accreting NS systems. Thus it would provide the opportunity to probe the evolution of magnetic inclination which may influence the evolution of pulsar structure, including the relaxation process in the differentially rotating core and the complex interplay between magnetism and relativistic hydrodynamics, with the Equation of State (EoS) also playing a role \citep{Lander2017MNRAS}. Future X-ray polarization observations on these accreting X-ray pulsars are expected to provide new independent measurements of the magnetic inclination and the geometry of X-ray binary, e.g., recent IXPE polarization measurement suggested the magnetic inclination of Cen X-3 to be $\sim 17^\circ$, corresponding to a near aligned rotator \citep{Tsygankov2022ApJ}. Therefore, the X-ray polarization combined with our methods will measure the magnetic inclination angle for more X-ray pulsars, probing the whole picture of magnetic structure evolution in accretion binaries, then should be interesting for NS physics.


\section*{Acknowledgements}

We are grateful to the referee for the useful suggestions to improve the manuscript.
This work is supported by the National Key Research and Development Program of China (Grants No. 2021YFA0718503), the NSFC (No. 12133007). This work has made use of data from the \textit{Insight}-HXMT mission, a project funded by the China National Space Administration (CNSA) and Chinese Academy of Sciences (CAS).

\section*{Data Availability}

Data used in this work are from the Institute of High Energy Physics, Chinese Academy of Sciences (IHEP-CAS) and have been publicly available for download from the \textit{Insight}-HXMT website http://hxmtweb.ihep.ac.cn/.



\bibliographystyle{mnras}
\bibliography{ref} 




\appendix

\section{Completed phase-resolved spectra}

\begin{figure*}
	\centering
	\includegraphics[width=1.0\textwidth]{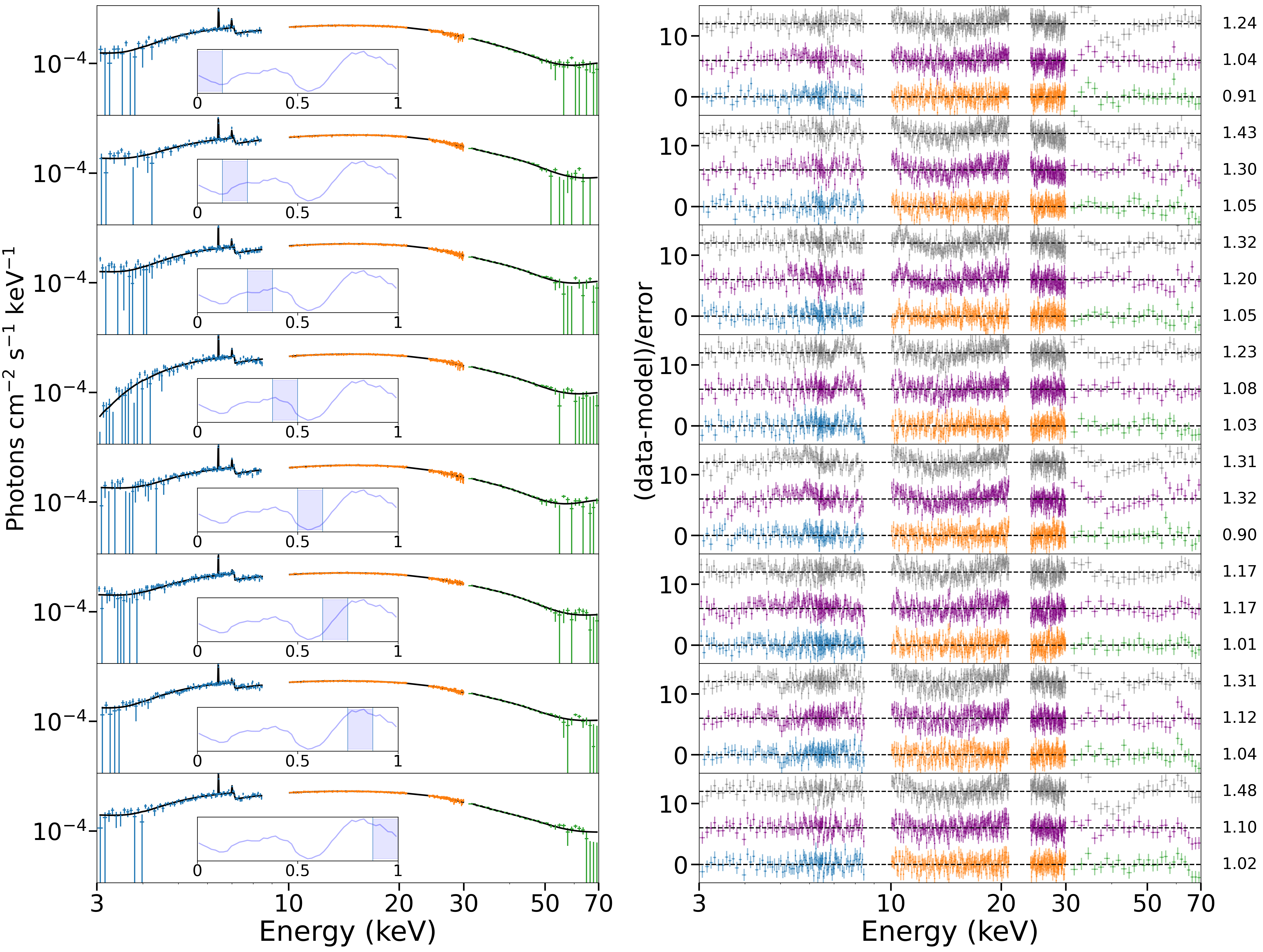}
	\caption{ Phase-resolved spectra derived in a super-critical observation. This is a completed version of Figure~\ref{fig:prspec} with all 8 phases. The ObsID is P010130900103 with $L = 4.2 \times 10^{37}$ erg$/$s. {\bf Left panels:} Fitting spectra with pulse profiles inserted, highlighting the corresponding pulse phases in each panel. The model used here is {\em Constant*TBabs*(TBpcf*gabs*gabs*Continuum+Gaussian+Gaussian)}. {\bf Right panels:} Residuals corresponding to the left panels with the same colors. Residuals from model with only one {\em gabs} are shown in each panel with purple lines, and their values are all increased by 6, while residuals from model with no absorption are shown in each panel with grey lines, and their values are all increased by 12. The reduced chi-square value is shown on the right side for each model. The broad band spectral shape is stable throughout all pulse phases. Almost all complications are due to the cyclotron (hard band) and neutral stellar wind (soft band) absorption.}
	\label{fig:prspec_all}
\end{figure*}

\begin{figure*}
	\centering
	\includegraphics[width=1.0\textwidth]{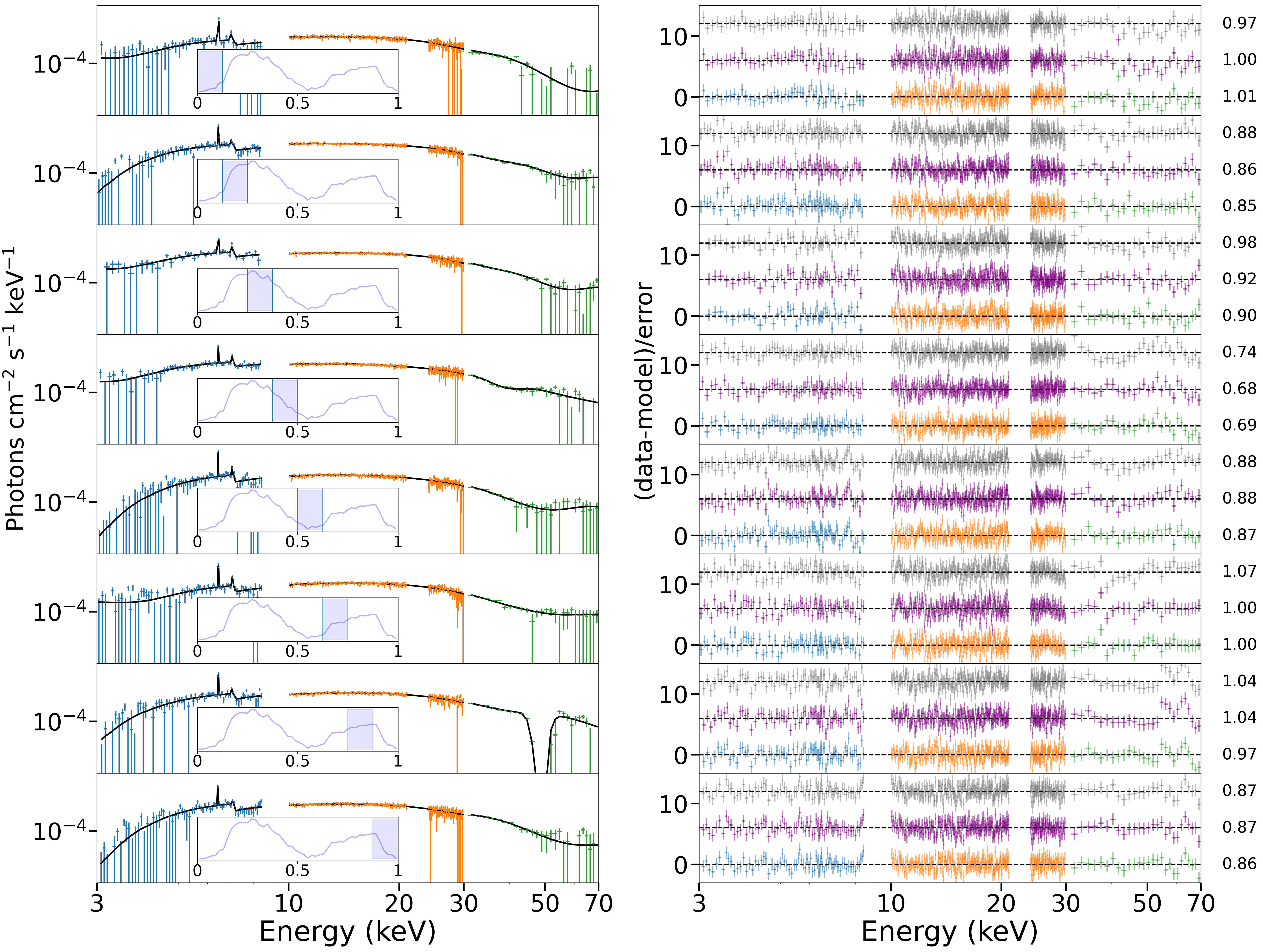}
	\caption{Phase-resolved spectra derived in a lower luminosity observation. This is a completed version of Figure~\ref{fig:prspec2} with all 8 phases. The ObsID is P010130900401 with $L = 6.6 \times 10^{36}$ erg$/$s. Elements are the same as Figure~\ref{fig:prspec_all}.}
	\label{fig:prspec2_all}
\end{figure*}



\bsp	
\label{lastpage}
\end{document}